\documentclass[12pt]{article}
\usepackage{graphicx}
\begin{document}

\begin{center}

{\Large \bf  Lorentz Group in \\Ray and Polarization Optics}

\vspace{10mm}

S. Ba{\c s}kal \footnote{electronic address:
baskal@newton.physics.metu.edu.tr}\\
Department of Physics, Middle East Technical University,
06531 Ankara, Turkey

\vspace{5mm}

Y. S. Kim \footnote{electronic address: yskim@physics.umd.edu}\\
Department of Physics, University of Maryland,\\
College Park, Maryland 20742, U.S.A.

\end{center}

\vspace{10mm}

\begin{abstract}
While the Lorentz group serves as the basic language for Einstein's
special theory of relativity, it is turning out to be the basic
mathematical instrument in optical sciences, particularly in ray
optics and polarization optics.  The beam transfer matrix, commonly
called the $ABCD$ matrix, is shown to be a two-by-two representation
of the Lorentz group applicable to the three-dimensional space-time
consisting of two space and one time dimensions.  The Jones matrix
applicable to polarization states turns out to be the two-by-two
representations of the Lorentz group applicable to the four-dimensional
space-time consisting of three space and one time dimensions.  The
four-by-four Mueller matrix applicable to the Stokes parameters as well
as the Poincar\'e sphere are both shown to be the representations of
the Lorentz group.
\end{abstract}

\vspace{15mm}

\noindent To be included as Chapter 10 in "Mathematical Optics: Classical,
Quantum and Imaging Methods" edited by Vasudevan Lakshminarayanan
(Taylor and Francis, New York).

\newpage

\section{Introduction}\label{intro}

One complex number contains two independent parameters.  One two-by-two
complex matrix with its four elements contains eight independent parameters.
The unit determinant condition reduces the number of independent parameters
to six.  These matrices form a group which is called
the two-dimensional special linear group and is denoted as $SL(2,c)$.
\par
This group has two important subgroups.  If we choose only Hermitian
matrices, this group is $SU(2)$ which is the two-dimensional rotation
group describing the rotations of electron spins.  Its correspondence
with the three-dimensional rotation group is well known.
\par
From the group $SL(2,c)$, we can choose matrices with real elements.
As in the case of the $SU(2)$ subgroup, these two-by-two real matrices
contain three independent parameters forming the $SL(2,r)$ or $Sp(2)$
subgroups.  In paraxial ray optics, reflections and refractions are governed
by such matrices and are called the $ABCD$ matrices~(Azzam et al. 1977,
Saleh et al. 2007).
Thus we shall collectively refer these real subgroups as the $ABCD$ group.
\par
In this report, we first discuss physical applications of this
three-parameter $ABCD$ group.  It is noted that the $ABCD$ matrix can
be decomposed into three easy-to-understand rotation and squeeze
matrices, which can then be rotated into a form having equal diagonal
elements and two independent parameters~(Ba\c{s}kal et al. 2009, Ba\c{s}kal et al. 2010).
We shall discuss how this process can be developed using optical instruments
for periodic systems, such as laser cavities and multilayer optics.

\par
While the group $SU(2)$ corresponds to $SO(3)$
consisting of three-by-three matrices, the $ABCD$ group corresponds to
the set of three-by-three matrices of Lorentz transformations applicable
to the three-dimensional space consisting of one time dimension and
two space dimensions. Lorentz transformations applicable to $z$ and $x$
directions and rotations around the $y$ axis are governed by this group,
which is called $SO(2,1)$.

\par
Consequently, this aspect of the $ABCD$ group allows us to study the essential
features of Wigner's little group~(Wigner 1939) dictating the internal
space-time symmetries of relativistic particles moving along the $z$
direction.  If we add the rotational degree of freedom around the
$z$ axis, the group can be extended to four-by-four matrices
applicable four-dimensional Minkowskian space consisting of three
space dimensions and one time dimension.
\par
Using the two-by-two matrix corresponding to the rotation around the
$z$ axis, we can extend the three-parameter $ABCD$ group to the
six-parameter $SL(2,c)$ group and thereafter to the six parameter
Lorentz group.  In this way we can move into polarization optics.
\par
The physics of light polarization has a long history~(Saleh et al. 2007,Brosseau 1998).
The basic mathematical instruments in this field are the two-by-two
matrix formalism for Jones vectors~(Jones 1941, 1947) and the four-by-four
Mueller~(Soleillet 1929, Mueller 1943) matrix formalism for the four
Stokes~(Stokes 1852) parameters.  For many years, it was known that
the set of matrices applicable to the two-component Jones vector has the
$SU(2)$ symmetry.  It is shown in this note that this $SU(2)$ symmetry can be
extended to that of $SL(2,c)$, if we take into account different
attenuation rates for the two different polarization
directions~(Opatrny et al.1993, Han et al. 1997, Ben-Aryeh 2005).

\par
In this report, we address the issue of unifying these two mathematical
devices into one mathematical formalism.  We show that
the Jones vector and the Stokes parameters are two-by-two and four-by-four
representations of the same Lorentz group.  Then the question is why
we need the Stokes parameters with a more complicated four-by-four
matrices.  The four-by-four formalism can tell the degree of coherence
between the two orthogonal polarization components.  However, the Lorentz
symmetry cannot change the decoherence parameter, in as much as the same
way as this symmetry cannot change the mass of a given particle.
In order to address this issue, we use the Poincar\'e sphere.
\par
The Poincar\'e sphere is a very useful and elegant graphical method
to represent the polarization state of light~(Poincar\'{e} 1982)
In the past it was regarded as a geometry of the three-dimensional sphere
with a fixed radius~(Born et al. 1980).  This sphere cannot accommodate the
symmetry of the Lorentz group.  Recently, it was noted by the present
authors that this three-dimensional geometry can be extended to the
four-dimensional Lorentz group~(Ba\c{s}kal et al. 2006).  It was noted
also that this extended Poincar\'e sphere, with its Lorentz symmetry, cannot
accommodate the variation of the decoherence parameter.  However, it
is possible to introduce two coupled Poincar\'e spheres using the
$SO(3,2)$ deSitter group~(Ba\c{s}kal et al. 2006).  This enlarged symmetry allows
variations of the decoherence parameter.
\par
The loss of coherence leads to an increase in the entropy of the system.
Indeed, the entropy can be computed from this extended Poincar\'e sphere.
By coupling two Poincar\'e spheres, we can transfer the entropy from one
sphere to the other.  This could serve as another example of Feynman's
rest of the universe~(Feynman 1972).  We note here that two coupled
harmonic oscillators could serve as an illustrative example of Feynman's
rest of the universe~(Han et al. 1999).

\par

In Sec.~\ref{gabcd}, we start with the six generators of the Lorentz group
and their closed set of commutation relations.  Starting from the
two-by-two Pauli matrices, there are six linearly independent two-by-two
matrices.  It is noted that the three of these matrices are real, and the
other three are purely imaginary.  The $ABCD$ matrix is generated by
those imaginary matrices.
\par
In Sec.~\ref{decomp}, it is shown that the optical $ABCD$ matrix can be
decomposed into a product of three convenient matrices which are rotation
and squeeze matrices.  In Sec.~\ref{periodic}, we use these properties to
deal with periodic systems in optics.
\par
In Sec.~\ref{sptime}, the mathematical language of the $ABCD$ matrix is
translated into that of the internal space-time symmetries of relativistic
elementary particles.  We note that the Wigner decomposition and the Bargmann
decomposition can be explained in terms of the decomposition properties
of the $ABCD$ matrix.

\par
In Sec.~\ref{jones}, it is shown that the two-by-two Jones vector formalism
is a representation of the Lorentz group.  We add a squeeze matrix to
the conventional $SU(2)$ formalism.  The symmetry group is $SL(2,c)$
corresponding to the full Lorentz group with six independent parameters.
From this, it is possible to construct a four-by-four representation of
the Lorentz group.  In Sec.~\ref{stokes}, we show that the Mueller
matrix applicable to the four Stokes parameters is the four-by-four
representation of the Lorentz group.

\par
Unlike the case of the two-by-two representation applicable to the Jones
vectors, the four-by-four representation can deal with the decoherence
problems.  On the other hand, the decoherence parameter remains as a
Lorentz-invariant parameter.  In order to deal with this problem, we
enlarge the symmetry group from the traditional Lorentz group of
$SO(3,1)$ to the deSitter group of $O(3.2)$.   We carry out this
operation by extending the concept of the Poincar\'e sphere.

\par
In Sec.~\ref{poins}, we first extend the traditional three-dimensional
sphere to a four-dimensional object, in order to take into account all
the symmetry contents of the Stokes parameters.  We then introduce two
Poincar\'s spheres coupled to each other through the symmetry of the
$O(3,2)$ deSitter group.  This allows the exchange of the decoherence
parameters between the two spheres.  Furthermore, this picture of the
two coupled Poincar\'e spheres constitute another example of Feynman's
rest of the universe~(Feynman 1972).

\section{Group of ABCD Matrices}\label{gabcd}
The Lorentz group is the group of four-by-four matrices applicable to
the four-dimensional Minkowskian space of $(x, y, z, t)$.  The group
is generated by three rotation generators $J_i$ and three boost
generators $K_i$, satisfying a closed set of commutation relations
\begin{equation}\label{commu01}
\left[J_i, J_j\right] = i\epsilon_{ijk} J_{k}, \quad
\left[J_i, K_j\right] = i\epsilon_{ijk} K_{k}, \quad
\left[K_i, K_j\right] = -i\epsilon_{ijk} J_{k} .
\end{equation}
These generators are all four-by-four matrices, and we shall give
their explicit expressions in later sections.
\par
Fortunately, there is a set of two-by-two matrices which satisfy the same
commutation relations.  We can write them as
\begin{equation}\label{gen01}
J_1 = \frac{1}{2}\pmatrix{0 & 1 \cr 1 & 0}, \qquad
J_2 = \frac{1}{2}\pmatrix{0 & -i \cr i & 0}, \qquad
J_3 = \frac{1}{2}\pmatrix{1 & 0 \cr 0 & -1} .
\end{equation}
They are the Pauli spin matrices which are very familiar to us. They
are Hermitian matrices.

\par
The boost generators $K_i$ can take the form of $iJ_i$, or explicitly
\begin{equation}\label{gen02}
K_1 = \frac{1}{2}\pmatrix{0 & i \cr i & 0}, \qquad
K_2 = \frac{1}{2}\pmatrix{0 & 1 \cr -1 & 0}, \qquad
K_3 = \frac{1}{2}\pmatrix{i & 0 \cr 0 & -i} .
\end{equation}
Unlike $J_i$, these matrices are anti-Hermitian.
\par
The group of two-by-two matrices generated by the above six two-by-two
matrices is called $SL(2,c)$.  Since the generators of this group share
the same set of commutation relations as the Lorentz group, they are
said to be locally isomorphic to each other.  In this paper, we shall
avoid this mathematical word, and simply say that $SL(2,c)$ is ``like''
the Lorentz group.
\par
There are a number of interesting subgroup of this $SL(2,c)$ group.
The three generators $J_i$ generates the $SU(2)$ subgroup which is like
the three-dimensional rotation group.  This aspect is well known.
Three $K_i$ alone do not form a closed set of commutation relations.

\par
Among the six generators, $J_2, K_1$, and $K_3$ are pure imaginary,
and they generate two-by-two matrices with real elements.  Furthermore
they satisfy the closed set of commutation relations
\begin{equation}\label{gen03}
\left[J_2, K_1\right] = -i K_3, \quad
\left[J_2, K_3\right] = i K_1, \quad
\left[K_1, K_3\right] = i J_2 .
\end{equation}
The group of two-by-two matrices generated by these three matrices is
called $Sp(2)$ or the two-dimensional symplectic group.  It is like the
Lorentz group applicable to the three-dimensional space of $(z, x, t)$.
However, this group constitutes the fundamental mathematical language
for the optical beam transfer matrix often called the $ABCD$ matrix.
\par

The $ABCD$ matrix is a two-by-two matrix with real elements, and
its determinant is one.  There are therefore three independent
parameters.  These elements are determined by optical materials
and how they are arranged.  The purpose of this note is to explore
its mathematical properties which can address more fundamental
issues in physics.
\par
First of all, the trace of this matrix could be less than two, equal
to two, or greater than two.  We are interested in what physical
conclusions we can derive from these numbers.
\par
In order to bring the $ABCD$ matrix to the form which will describe
the particle symmetries, we should first transform it into the
equi-diagonal form where the two diagonal elements are equal to
each other~(Ba\c{s}kal et al. 2009, Ba\c{s}kal et al. 2010).
We can achieve this goal by a similarity transformation with a rotation
matrix.  Thus, this rotation angle becomes one of the three independent
parameters, and the equi-diagonal $ABCD$ matrix has two independent
parameters.
\par
We shall call this equi-diagonal matrix the core of the $ABCD$ matrix,
and use the notation $[ABCD]$.  This matrix is not always diagonalized.
This creates non-trivial problem.  We shall examine how optical
devices, especially periodic systems, can lead us to a better understanding
of the problem.  For this purpose, we discuss laser cavities and
multilayer systems in detail.
\par
If the trace is less than two, the core can be written as
\begin{equation}\label{core01}
[ABCD] =\pmatrix{\cos(\gamma/2) & - e^{\eta} \sin(\gamma/2) \cr
  e^{-\eta} \sin(\gamma/2)  & \cos(\gamma/2)} .
\end{equation}
The diagonal elements are equal and smaller than one.
\par
If the trace is greater than two, the core takes the form
\begin{equation}\label{core02}
[ABCD] = \pmatrix{\cosh(\gamma/2) & e^{\eta} \sinh(\gamma/2) \cr
  e^{-\eta} \sinh(\gamma/2)  & \cosh(\gamma/2)} .
\end{equation}
Here again the diagonal elements are equal, but they are greater than one.
\par
If the trace is equal to two, the $[ABCD]$ matrix becomes
\begin{equation}\label{core03}
[ABCD] = \pmatrix{1 &  -\gamma \cr 0  & 1 } .
\end{equation}
This matrix also has the same diagonal element, and they are equal
to one.

\par
The triangular matrix of Eq.(\ref{core03}) cannot be diagonalized.
As for the matrices of Eq.(\ref{core01}) and Eq.(\ref{core02}), they
can be diagonalized, but not by rotation alone.  These mathematical
subtleties are not well known.  The purpose of this report is
to show how much physics we can understand by studying this
mathematical subtleties.

\par
The mathematics of group theory allows us to write down a four-by-four
Lorentz-transformation matrix for every two-by-two matrix discussed in
this paper.  In this way, the three matrices given in Eq.(\ref{core01}),
Eq.(\ref{core02}), and Eq.(\ref{core03}) lead to the internal
space-time symmetries of elementary particles.  They respectively
correspond to the symmetries of massive, imaginary-mass, and massless
particles, respectively~(Wigner 1939, Kim et al. 1986).

\section{Decomposition of the ABCD Matrix}\label{decomp}
We are interested in writing the three different forms of
the core matrix in one expression.
\begin{equation}\label{core10}
[ABCD] = \exp{\left\{\frac{1}{2}
   \pmatrix{0 & - x - y  \cr x - y & 0}\right\}} ,
\end{equation}
where the parameters $x$ and $y$ are determined by the optical
materials and how they are arranged.  The exponent of this
matrix is
\begin{equation}
\frac{1}{2} \pmatrix{0 & - x - y  \cr x - y & 0} .
\end{equation}
\par

If $x > y $, the exponent becomes
\begin{equation}
\frac{\gamma}{2} \pmatrix{0 & -\exp{(\eta)}  \cr \exp{(-\eta)} & 0} ,
\end{equation}
which leads to the core matrix of Eq.(\ref{core01}) with
\begin{eqnarray}\label{core11}
&{}& \gamma = \sqrt{x^2 - y^2}, \nonumber\\[1ex]
&{}& e^{\eta} = \sqrt{\frac{x + y}{x - y}} .
\end{eqnarray}
The core matrix $[ABCD]$ can be written as a similarity transformation
\begin{equation}\label{core16}
[ABCD] = B(\eta) R(\theta) B(-\eta)
\end{equation}
with
\begin{eqnarray}\label{core26}
&{}& B(\eta) = \pmatrix{e^{\eta/2} & 0 \cr 0 & e^{-\eta/2}},  \nonumber \\[2ex]
&{}& R(\theta) = \pmatrix{\cos(\theta/2) & -\sin(\theta/2)
     \cr \sin(\theta/2) & \cos(\theta/2)} ,
\end{eqnarray}
where $\gamma$ is now replaced by the rotation angle $\theta$.
$R(\theta)$ is a rotation matrix, and $B(\eta)$ is a squeeze matrix.

\par
If $x < y $, the exponent becomes
\begin{equation}
\frac{\gamma}{2}
   \pmatrix{0 & -\exp{(\eta)}  \cr -\exp{(-\eta)} & 0},
\end{equation}
leading to the core matrix of Eq.(\ref{core02}), with
\begin{eqnarray}\label{core12}
&{}& \gamma = \sqrt{y^2 - x^2}, \nonumber\\[1ex]
&{}& e^{\eta} = \sqrt{\frac{x + y}{y - x}} .
\end{eqnarray}
The $[ABCD]$ matrix can now be decomposed into a similarity transformation
\begin{equation}\label{core15}
[ABCD] = B(\eta) S(-\lambda) B(-\eta) ,
\end{equation}
with
\begin{equation}\label{core36}
S(\lambda) = \pmatrix{\cosh(\lambda/2) & \sinh(\lambda/2)
     \cr \sin(\lambda/2) & \cosh(\lambda/2)} ,
\end{equation}
where $\gamma$ is replaced by the boost parameter $\lambda$.  The
matrix $B(\eta)$ takes the diagonal form given in Eq.(\ref{core16})
with $\eta$ defined in Eq.(\ref{core12}).  $S(\lambda)$ is a squeeze
matrix.
\par
If $x = y$, the exponent becomes
\begin{equation}\label{core55}
  \pmatrix{1 &  - x \cr 0  & 1 } ,
\end{equation}
with $x = y = \gamma$.
\par
We now have combined three different expressions for the core of the $ABCD$
matrix into one exponential form of Eq.(\ref{core10}). This form can be
decomposed into three matrices constituting a similarity transformation.

\par
There is another form of decomposition known as the Bargmann
decomposition~(Bargmann 1947), which states that the core of the $ABCD$ matrix
can be written as
\begin{equation}\label{barg11}
[ABCD] = R(\alpha) S(-2\chi) R(\alpha) ,
\end{equation}
where the forms of the rotation matrix $R$ and the squeeze matrix $S$ are
given as in Eq.(\ref{core26}) and Eq.(\ref{core36}) respectively.  If we carry
out the matrix multiplication, the $[ABCD]$ matrix becomes
\begin{equation}\label{barg22}
\pmatrix{(\cosh\chi)\cos\alpha  &
       -\sinh\chi - (\cosh\chi)\sin\alpha  \cr
      -\sinh\chi + (\cosh\chi)\sin\alpha  &
  (\cosh\chi)\cos\alpha } .
\end{equation}
This matrix also has two independent parameters $\alpha$ and $\chi$.
We can write these parameters in terms of $\theta$ and $\eta$ by
comparing the matrix elements.  For instance, if $x > y$, the
diagonal elements lead to
\begin{equation}
\cos(\theta/2) = (\cosh\chi) \cos\alpha .
\end{equation}
The off-diagonal elements lead to
\begin{equation}
e^{2\eta} = \frac{(\cosh\chi)\sin\alpha + \sinh\chi}
           {(\cosh\chi)\sin\alpha - \sinh\chi}.
\end{equation}

\par

As for physical applications, let us consider periodic systems,
such as laser cavities and multilayer systems.  The exponential
form given in Eq.(\ref{core10}) tells us that it is a matter of
replacing the $\theta$ parameter by $N \theta$ for N repeated
applications~(Ba\c{s}kal et al. 2010).  Let us see some examples.
\par

\section{Periodic Systems in Ray optics}\label{periodic}

Let us summarize the content of Sec.~\ref{decomp}.  First of all, the optical
$ABCD$ matrix can be brought to an equi-diagonal form by a similarity
transformation, and we call this equi-diagonal matrix the core of $ABCD$,
and use the notation $[ABCD]$.
Thus,
\begin{equation}
(ABCD) = T~[ABCD]~T^{-1} ,
\end{equation}
where we use $(ABCD)$ for the original $ABCD$ matrix.  The transformation
matrix can be a rotation or a triangular matrix.  In addition, the core
matrix can be written as a Wigner decomposition of the form
\begin{equation}
[ABCD] = B(\eta)~W(\gamma)~B(-\eta),
\end{equation}
which is another similarity transformation with $B(\eta)$ given in
Eq.(\ref{core26}).  The $W(\gamma)$ is one of the three matrices
\begin{equation}
\pmatrix{\cos(\gamma/2) & -\sin(\gamma/2) \cr \sin(\gamma/2) &
        \cosh(\gamma/2)}, \quad
\pmatrix{\cosh(\gamma/2) & -\sinh(\gamma/2) \cr -\sinh(\gamma/2) &
          \cosh(\gamma/2)}, \quad
\pmatrix{1  & -\gamma \cr 0 & 1},
\end{equation}
and
\begin{equation}\label{sim66}
(ABCD) = [T~B(\eta)]~W(\gamma)~[T~B(\eta)]^{-1} .
\end{equation}
\par
For repeated applications of the $ABCD$ matrix we need an expression
of the form $(ABCD)^N$.
Thanks to this form of similarity transformation, $(ABCD)^N$
is now simplified as
\begin{equation}
(ABCD)^N = [T~B(\eta)]~W(N\gamma)~[T~B(\eta)]^{-1} .
\end{equation}
Thus the problem of periodic systems in optics is to find the core matrix
$[ABCD]$.

\subsection{Laser Cavities}\label{lacav}

As the first example, let us consider
the laser cavity consisting of two identical concave mirrors separated
by a distance $d$.  Then the $ABCD$ matrix
for a round trip of one beam is
\begin{equation}\label{abcd}
  \pmatrix{1 & 0 \cr -2/R & 1}
  \pmatrix{1 & d \cr 0 & 1}
  \pmatrix{1 & 0 \cr -2/R & 1} \pmatrix{1 & d \cr 0 & 1},
\end{equation}
where
the matrices
\begin{equation}\label{rad}
  \pmatrix{1 & 0 \cr -2/R & 1} ,  \quad
  \pmatrix{1 & d \cr 0 & 1}
\end{equation}
are the mirror and translation matrices respectively.  The parameters $R$
and $d$ are the radius of the mirror and the mirror separation respectively.
This form is quite familiar to us from the laser
literature~(Yariv 1975, Haus 1984, Hawkes et al. 1995).
\par
However, the main issue here is how to alleviate the problem of taking the $Nth$
power of chains of matrices which corresponds to the repetition of this process
for many times.  This can be achieved when the matrices in Eq.(\ref{abcd})
can be brought to an equi-diagonal form and eventually to a form of the
Wigner decomposition.  Thus, we are interested in finding the core of
Eq.(\ref{abcd}).  For his purpose,  we rewrite the matrix of
Eq.(\ref{abcd}) as
\begin{eqnarray}\label{abcd3}
&{}&  \pmatrix{1 & -d/2 \cr 0 & 1}
  \pmatrix{1 & d/2 \cr 0 & 1}
  \pmatrix{1 & 0 \cr -2/R & 1}
  \pmatrix{1 & d/2 \cr 0 & 1}^{2} \nonumber\\[1ex]
&{}&  \times
  \pmatrix{1 & 0 \cr -2/R & 1}
  \pmatrix{1 & d/2 \cr 0 & 1}
  \pmatrix{1 & d/2 \cr 0 & 1} .
\end{eqnarray}

\par
In this way, we translate the system by $-d/2$ using a translation
matrix given in Eq.(\ref{rad}), and  write the
$ABCD$ matrix of Eq.(\ref{abcd}) as
\begin{equation}
\pmatrix{1 & -d/2 \cr 0 & 1}
\left[\pmatrix{1 - d/R &   d - d^{2}/2R  \cr -2/R & 1 - d/R}\right]^{2}
\pmatrix{1 & d/2 \cr 0 & 1}.
\end{equation}
We are thus led to concentrate on the matrix in the middle
\begin{equation}
 \pmatrix{1 - d/R &   d - d^{2}/2R  \cr -2/R & 1 - d/R},
\end{equation}
which can be written as
\begin{equation}
 \pmatrix{\sqrt{d} & 0  \cr 0 & 1/\sqrt{d}}
  \pmatrix{1 - d/R &   1 - d/2R  \cr -2d/R & 1 - d/R}
  \pmatrix{1/\sqrt{d} & 0  \cr 0 & \sqrt{d}} .
\end{equation}
\par
It is then possible to decompose the $ABCD$ matrix into
\begin{equation}
E~C^{2}~E^{-1} ,
\end{equation}
with
\begin{eqnarray}\label{escort}
&{}&  C = \pmatrix{1 - d/R &  1 - d/2R  \cr
         -2d/R & 1 - d/R},  \nonumber \\[2mm]
&{}&  E = \pmatrix{1 & -d/2 \cr 0 & 1}
\pmatrix{\sqrt{d} & 0 \cr 0 & 1/\sqrt{d}} .
\end{eqnarray}
The $C$ matrix now contains only dimensionless numbers, and it
can be written as
\begin{equation}
C = \pmatrix{\cos(\gamma/2) & e^{\eta}\sin(\gamma/2)  \cr
    - e^{-\eta} \sin(\gamma/2)  & \cos(\gamma/2)} ,
\end{equation}
with
\begin{eqnarray}
&{}&\cos(\gamma/2) = 1 - \frac{d}{R},   \nonumber\\[2ex]
&{}& e^\eta = \sqrt{\frac{2R - d}{4d}} .
\end{eqnarray}
Here both $d$ and $R$ are positive, and the restriction on them is
that $d$ be smaller than $2R$.  This is the stability condition
frequently mentioned in the literature~(Haus 1984, Hawkes et al. 1995).
\par

Thus, the $[ABCD]$ core matrix is $C^2$, and takes the form
\begin{equation}
[ABCD]  = \pmatrix{\cos(\gamma) & e^{\eta}\sin(\gamma)  \cr
    - e^{-\eta} \sin(\gamma)  & \cos(\gamma)} ,
\end{equation}
and the similarity transformation which connects this core matrix
with the original $ABCD$ matrix of Eq.(\ref{abcd}) is $E$ given
in Eq.(\ref{escort}).

\subsection{Multilayer Optics}\label{multi}
We consider an optical beam going through a periodic medium with
two different refractive indices.   If the beam traveling in the
first medium hits the second medium, it is partially transmitted
and partially reflected.  In order to maintain the continuity of
the Poynting vector, we define the electric fields as
\begin{eqnarray}
E_{1}^{(\pm)} = \frac{1}{\sqrt{n_1}}
       \exp{\left({\pm}ik_1 z  - \omega t \right)} \nonumber\\[1ex]
E_{2}^{(\pm)} = \frac{1}{\sqrt{n_2}}
           \exp{\left({\pm}ik_2 z  - \omega t \right)}
\end{eqnarray}
for the optical beams in the first and second media respectively.
The superscript $(+)$ and $(-)$ are for the incoming and reflected
rays respectively.
\par
These two optical rays are related by the two-by-two $ABCD$ matrix,
according to
\begin{equation}\label{smat}
\pmatrix{E_{2}^{(+)} \cr E_{2}^{(-)}} =
\pmatrix{A & B \cr C & D}
\pmatrix{E_{1}^{(+)} \cr E_{1}^{(-)}} .
\end{equation}
Of course the elements of this matrix are determined by
transmission coefficients as well as the phase shifts the beams
experience while going through the media~(Azzam et al. 1977, Georgieva et al. 2001).
\par
When the beam goes through the first medium to the second, we may use the
the boundary matrix given in~(Azzam et al. 1977) and
in~(Monzon et al. 2000, Monzon et al.2002).  In terms
of the refractive indexes $n_1$ and $n_2$, this matrix is
\begin{equation}\label{boundary11}
Q(\sigma) = \pmatrix{\cosh(\sigma/2) & \sinh(\sigma/2) \cr
              \sinh(\sigma/2) & \cosh(\sigma/2) } ,
\end{equation}
where one can write the $\sigma$ parameter as
\begin{equation}
\cosh\left(\frac{\sigma}{2}\right) = \frac{n_1 + n_2}{2\sqrt{n_1 n_2}}, \qquad
\sinh\left(\frac{\sigma}{2}\right) = \frac{n_1 - n_2}{2\sqrt{n_1 n_2}}.
\end{equation}
The boundary matrix for the beam going from the second medium
should be $Q(-\sigma)$.
\par
In addition, we have to consider the phase shifts the beams have to
go through.  When the beam goes trough the first media, we can use
the phase-shift matrix
\begin{equation}
P\left(\delta_1\right) =
\pmatrix{e^{-i\delta_1/2} &  0 \cr 0 & e^{i\delta_1/2} }   ,
\end{equation}
and a similar expression for $P\left(\delta_2\right)$ for the second
medium.  The phase shift $\delta$ is determined by the wave number and
the thickness of the medium.
\par
We are thus led to consider one complete cycle starting from the midpoint
of the second medium, and write
\begin{equation} \label{chain11}
 P\left(\delta_2/2\right) Q(\sigma) P\left(\delta_1\right) Q(-\sigma)
  P\left(\delta_2/2\right) .
\end{equation}
\par
There are two questions in regards to the above matrix multiplication.
One is whether each matrix in this expression can be converted into
a matrix with real elements and the other is whether the result of this
matrix multiplication accommodates Wigner and Bargmann decompositions.
In order to answer the first question, let us consider the similarity
transformation
\begin{equation}\label{conju66}
C_1~P(\delta) Q(\sigma)~C_1^{-1} ,
\end{equation}
with
\begin{equation}
C_{1} =  \frac{1}{\sqrt{2}} \pmatrix{1 & i \cr i & 1} .
\end{equation}
This transformation leads to
\begin{equation} \label{conju77}
  R(\delta) Q(\sigma) ,
\end{equation}
where
\begin{equation}
R(\delta) = \pmatrix{\cos(\delta/2) & -\sin(\delta/2) \cr
                     \sin(\delta/2) & \cos(\delta/2) } .
\end{equation}
This notation is consistent with the rotation matrices
used in Sec.~\ref{decomp}.

\par
Let us make another similarity transformation with
\begin{equation}
C_{2} =  \frac{1}{\sqrt{2}} \pmatrix{1 & 1 \cr -1 & 1} .
\end{equation}
This changes $Q(\sigma)$ into $B(\sigma)$ without changing $R(\delta)$,
where
\begin{equation}
B(\sigma) = \pmatrix{e^{\sigma/2} & 0 \cr 0 & e^{-\sigma/2}},
\end{equation}
again consistent with the $B(\eta)$ matrix used in
Sec.~\ref{decomp}.

 \par

Thus the net similarity transformation matrix is~(Georgieva et al. 2001)
\begin{equation}\label{ccc}
C = C_{2}C_{1} = \frac{1}{\sqrt{2}} \pmatrix{e^{i\pi/4} &  e^{i\pi/4}
\cr -e^{-i\pi/4} & e^{-i\pi/4}}.
\end{equation}

\par

If we apply this similarity transformation to the long matrix chain of
Eq.(\ref{chain11}), it becomes another chain
\begin{equation} \label{chain22}
M = R\left(\delta_2/2\right) B(\sigma) R\left(\delta_1\right)
    B(-\sigma) R\left(\delta_2/2\right) ,
\end{equation}
where all the matrices are real.
\par

Let us now address the main question of whether this matrix chain can
be brought to one equi-diagonal matrix.  We note first that the
three middle matrices can be written in a familiar form
\begin{eqnarray} \label{chain33}
&{}& \hspace{-5mm} M =  B(\sigma) R\left(\delta_1\right) B(-\sigma) \nonumber \\[2ex]
&{}& = \pmatrix{\cos(\delta_1/2) & -e^{\sigma}\sin(\delta_1/2) \cr
 e^{-\sigma}\sin(\delta_1/2)& \cos(\delta_1/2)}
\end{eqnarray}
However, due to the rotation matrix $R\left(\delta_2/2\right)$ at the
beginning and at the end of Eq.(\ref{chain22}), it is not clear whether the
entire chain can be written as a similarity transformation.
\par
In order to resolve this issue, let us write Eq.(\ref{chain33}) as a Bargmann
decomposition
\begin{equation}\label{barg66}
R(\alpha) S(-2\chi) R(\alpha) ,
\end{equation}
with its explicit expression given in Eq.(\ref{barg22}).  The parameters
$\alpha$ and $\chi$ are related to $\sigma$ and $\delta_1$ by
\begin{eqnarray}
&{}& \cos(\delta_1/2) = (\cosh\chi) \cos\alpha ,  \nonumber \\[2ex]
&{}& e^{2\sigma} = \frac{(\cosh\chi)\sin\alpha + \sinh\chi}
           {(\cosh\chi)\sin\alpha - \sinh\chi} .
\end{eqnarray}
\par
It is now clear that the entire chain of Eq.(\ref{chain11}) can be written
as another Bargmann decomposition
\begin{equation}\label{barg77}
M = R(\alpha + \delta_2/2) S(-2\chi) R(\alpha + \delta_2/2) .
\end{equation}
Finally, this expression can be converted to a Wigner
decomposition~(Georgieva et al. 2003)
\begin{equation}\label{barg88}
M = B(\eta) R(\theta) B(-\eta),
\end{equation}
with
\begin{eqnarray}
&{}& \cos(\theta/2) = (\cosh\chi) \cos(\alpha + \delta_2/2),  \nonumber \\[2ex]
&{}& e^{2\eta} = \frac{(\cosh\chi)\sin(\alpha + \delta_2/2) + \sinh\chi}
           {(\cosh\chi)\sin(\alpha + \delta_2/2) - \sinh\chi} .
\end{eqnarray}
The decomposition of Eq.(\ref{barg88}) allows us to deal with the periodic
system of multilayers.  For repeated application of $M$, we can now write
\begin{equation}\label{repeat33}
M^N = B(\eta) R(N\theta) B(-\eta).
\end{equation}

\section{Space-time Symmetries}\label{sptime}
In Sec.~\ref{decomp}, we have seen that the two-by-two matrices can
provide a very powerful language for optical systems.  However,
this language is not restricted to the two-dimensional world.  It
can be translated into the four-dimensional world of Einstein's
special relativity where Lorentz transformations play the central
role.

\par
In mathematics, the group of two-by-two unimodular matrices is called
$SL(2,c)$.  The group of four-dimensional matrices performing Lorentz
transformations on the Minkowskian four-vector $(t, z, x, y)$ is called
the $SO(3,1)$ Lorentz group.  The group $SL(2,c)$ has six generators,
so does the $SO(3,1)$, but the corresponding matrices are two-by-two
and four-by-four, respectively.  Their generators satisfy the same
set of commutation relations as in Eq.(\ref{commu01}). This correspondence
is called the local isomorphism between the $SL(2,c)$ and $SO(3,1)$ groups.
The four-dimensional generators are
\begin{equation}\label{4gen01}
J_{1} = \pmatrix{0 & 0 & 0 & 0 \cr 0 & 0 & 0 &  i
\cr 0 & 0 & 0 & 0 \cr 0 & -i & 0 & 0 }, \quad
J_{2} = \pmatrix{0 & 0 & 0 & 0 \cr 0 & 0 & -i & 0
\cr 0 & i & 0 & 0 \cr 0 & 0 & 0 & 0 },   \quad
J_{3} = \pmatrix{0 & 0 & 0 & 0 \cr 0 & 0 & 0 & 0
\cr 0 & 0 & 0 & -i \cr 0 & 0 & i & 0 } ,
\end{equation}
and
\begin{equation}\label{4gen02}
K_{1} = \pmatrix{0 & 0 & i & 0 \cr 0 & 0 & 0 & 0
   \cr i & 0 & 0 & 0 \cr 0 & 0 & 0 & 0 }, \quad
K_{2} = \pmatrix{0 & 0 & 0 & i \cr 0 & 0 & 0 & 0
    \cr i & 0 & 0 & 0 \cr 0 & 0 & 0 & 0 }, \quad
K_{3} = \pmatrix{0 & i & 0 & 0 \cr i & 0 & 0 & 0
    \cr 0 & 0 & 0 & 0 \cr 0 & 0 & 0 & 0 } .
\end{equation}
\par
This mathematical property allows us to explain events in Einstein's
Lorentz-covariant world in terms of what we observe in optics
laboratories.

\subsection{Two-by-two and four-by-four representations of the Lorentz group}

The content of this correspondence is somewhat complicated, but for the
present purpose, we can start with the Minkowskian four-vector
$(t, z, x, y)$ written as
\begin{equation} \label{slc01}
X = \pmatrix{t + z & x - iy \cr x + iy & t - z} ,
\end{equation}
whose determinant
\begin{equation}
   t^2 - z^2 - x^2 - y^2
\end{equation}
is left invariant under Lorentz transformations.
Now consider a transformation
\begin{equation}\label{slc02}
X' = G~X~G^{\dagger} ,
\end{equation}
where $G$ a unimodular matrix whose determinant is one.  Let us write
this matrix as
\begin{equation}\label{alphabeta}
G = \pmatrix{\alpha & \beta \cr \gamma & \delta} ,
\end{equation}
where the elements can be complex numbers.  If the determinant of this
matrix is one, there are only six independent parameters.  Thus this
matrix can be generated by the six generators given in Eq.(\ref{gen01})
and Eq.(\ref{gen02}).
\par
The transformation of Eq.(\ref{slc02}) can be explicitly written as
\begin{equation}\label{slc04}
\pmatrix{t' + z' & x' - iy' \cr x' + iy' & t' - z'} =
\pmatrix{\alpha & \beta \cr \gamma & \delta}
\pmatrix{t + z & x - iy \cr x + iy & t - z}
\pmatrix{\alpha^* & \gamma^* \cr \beta^* & \delta^*} .
\end{equation}
We can now translate this formula into
\begin{equation}\label{translate}
\pmatrix{t' + z' \cr x' - iy' \cr x' + iy' \cr  t' - z'} =
\pmatrix{\alpha\alpha^{*} & \alpha \beta^{*} &
\beta\alpha^{*} & \beta \beta^{*} \cr
\alpha \gamma^{*} & \alpha \delta^{*} &
\beta \gamma^{*} & \beta \delta^{*} \cr
\gamma \alpha^{*}  & \gamma \beta^{*} &
\delta \alpha^{*} & \delta \beta^{*} \cr
\gamma \gamma^{*} & \gamma \delta^{*} &
\delta \gamma^{*} & \delta \delta^{*}}
\pmatrix{t + z \cr x - iy  \cr x + iy \cr t -z} .
\end{equation}
It can be seen that the above transformation matrix can be
expressed in the form of a Kronecker product as
\begin{equation}\label{kronecker}
G \otimes G^{*}
\end{equation}
where
\begin{equation}
G^{*}=\pmatrix{\alpha^{*} & \beta^{*} \cr \gamma^{*} & \delta^{*}}.
\end{equation}
Then the components of $X'$ are related to $(t', z', x', y')$ as
\begin{equation}
\pmatrix{t' \cr z' \cr x' \cr y'} = \frac{1}{2}
\pmatrix{1 & 1 & 0 & 0 \cr 1 & -1 & 0 & 0 \cr
 0 & 0 & 1 & 1 \cr 0 & 0 & i & -i}
\pmatrix{t' + z' \cr t' - z' \cr x' - iy' \cr x' + iy'} .
\end{equation}

\par
Likewise, the two-by-two matrix for the four-momentum of the particle takes the form
\begin{equation}\label{slc05}
P = \pmatrix{p_0 + p_z &  p_x - ip_y \cr p_x + ip_y &  p_0 - p_z}
\end{equation}
with $p_0 = \sqrt{m^2 + p_z^2 + p_x^2 + p_z^2}.$
The transformation of this matrix takes the same form as that of
the space-time four-vector given in Eq.(\ref{slc02}) and Eq.(\ref{slc04}).

\subsection{Internal Space-time Symmetries of Elementary Particles}

These properties are applicable to many other branches of physics.
For instance, one of the  persisting problem is the internal space-time
symmetry of elementary particles in Einstein's Lorentz-covariant world.
The mathematics of group theory
allows us to translate the rotation and squeeze matrices of
Eq.(\ref{core26}) and Eq.(\ref{core36}) into the following four-by-four
matrices respectively.
\begin{eqnarray}\label{4by4}
&{}& R(\theta) = \pmatrix{1 & 0 & 0 & 0 \cr
  0 & \cos\theta & -\sin\theta & 0 \cr
  0 & \sin\theta & \cos\theta &  0 \cr
  0 & 0 & 0 & 1} ,          \quad
  S(\lambda) = \pmatrix{\cosh\lambda & 0 &  \sinh\lambda & 0 \cr
  0 & 1 & 0 & 0 \cr
  \sinh\lambda & 0 & \cosh\lambda &  0 \cr
  0 & 0 & 0 & 1} , \nonumber\\[1ex]
&{}& B(\eta) = \pmatrix{\cosh\eta & \sinh\eta & 0 & 0 \cr
  \sinh\eta & \cosh\eta & 0 & 0 \cr
  0 & 0 & 1 &  0 \cr
  0 & 0 &  0 & 1} .
\end{eqnarray}

\par
They are applicable to the Minkowskian four-vector $(x, y, z, t)$.
The $R(\theta)$ matrix performs a rotation around the $y$ axis, and
$S(\lambda)$ is for Lorentz boosts along the $x$ axis.
The $B(\eta)$ matrix boosts the system along the $z$ direction.

\par
Together with a rotation matrix around $z$ axis~(Han et al. 1986)
\begin{equation}\label{z44}
Z(\phi) = \pmatrix{1 & 0 & 0 & 0\cr
   0 & 1 & 0 & 0 \cr
  0 & 0 & \cos\phi &  -\sin\phi \cr
  0 & 0 & \sin\phi & \cos\phi} ,
\end{equation}
they constitute Wigner's little groups dictating internal space-time
symmetries of massive and imaginary-mass particles~(Wigner 1939).
The triangular matrix of Eq.(\ref{core03}) leads to the little group
for massless particles.  The little groups are the subgroups of the
Lorentz group whose transformations leave the four-momentum of a
relativistic particle invariant.
\par
It is possible to compute the two-by-two equivalent of the above $Z(\phi)$
matrix using the relation given in Eq.(\ref{slc02}).  It takes the form
\begin{equation}\label{z22}
Z(\phi) = \pmatrix{e^{i\phi/2} & 0 \cr 0 & e^{-i\phi/2}} .
\end{equation}
This matrix contains complex elements.  This is the reason why it is
not mentioned in our discussions of the $ABCD$ matrix.  This rotation matrix
will play an important role in polarization optics which will be discussed
in Sec.~\ref{jones} and Sec.~\ref{stokes}.

\par

Let us go back to Eq.(\ref{core01}) which, according to Eq.(\ref{core16}),
can be decomposed to a similarity transformation
\begin{equation}\label{sim44}
   W(\eta, \theta) = B(\eta) R(\theta) B(-\eta).
\end{equation}
We can write this decomposition with the four-by-four matrices given
in Eq.(\ref{4by4}).
\par
Let us then consider a massive particle moving along the $z$
direction with the velocity parameter $v/c = \tanh\eta$, and its
four-momentum
\begin{equation} \label{mm1}
 (m\cosh\eta, m\sinh\eta, 0, 0),
\end{equation}
where $m$ is the mass of the particle.
\par

\begin{figure}[thb]
\centerline{\includegraphics[scale=0.29]{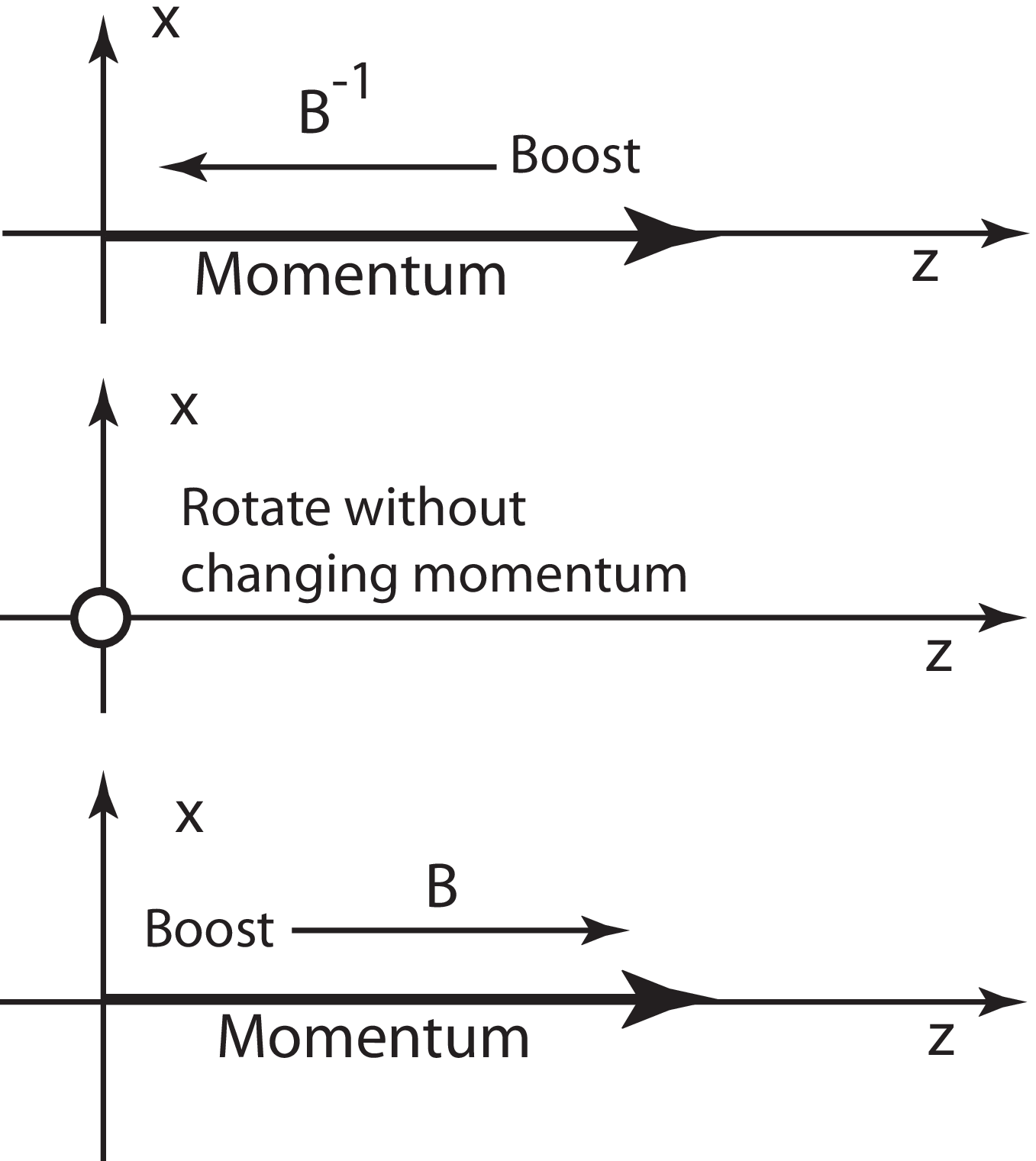}
\hspace{3mm} \includegraphics[scale=0.22]{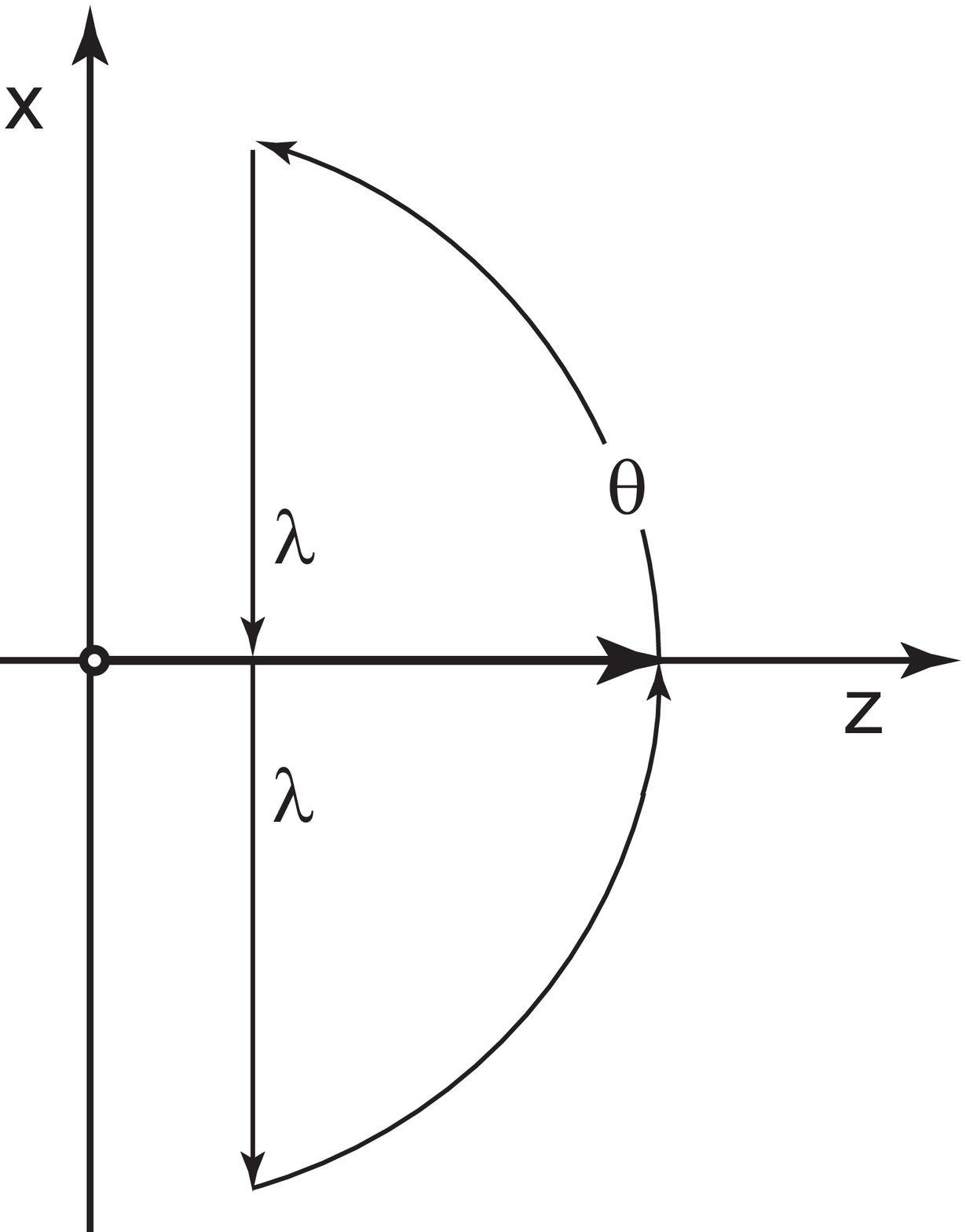}}
\caption{Wigner decomposition (left) and Bargmann decomposition (right).
These figures illustrate momentum preserving transformations.  In the
Wigner transformation, a massive particle is brought to its rest frame.
It can be rotated while the momentum remains the same.  This particle
is then boosted back to the frame with gaining original momentum.  In the
Bargmann decomposition, the momentum is rotated, boosted, and rotated
to its original state.}\label{bargwig}
\end{figure}

\par

We can boost this particle using the boost matrix $B(-\eta)$, which is
the inverse of the four-by-four matrix  given in Eq.(\ref{4by4}).   The
particle becomes at rest, with its four-momentum
\begin{equation}
 (m, 0, 0, 0),
\end{equation}
and with zero velocity.  The rotation matrix
$R(\theta)$ rotates this particle without changing its momentum.  During
this process, the particle changes the direction of its spin.  Finally,
$B(\eta)$ boosts the particle and restores its momentum.  In this way,
the four-by-four expression for Eq.(\ref{core16})
changes the internal space-time structure of the particle.
\par
One key question from this table is what happens to the $O(3)$-like
little group when the particle momentum becomes infinity or its mass
becomes zero.
The question is whether the little group for a massive particles becomes
that of a massless particle.  The answer to this question is Yes, but
this issue had a stormy history before this definitive answer~(Kim et al. 1990).
Indeed, when $\eta$ becomes infinity, the four-by-four form of Eq(\ref{sim44})
becomes
\begin{equation}\label{gamma44}
\pmatrix{1 + \gamma^2/2 & \gamma^2/2 & \gamma & 0 \cr
 \gamma^2/2 & 1 + \gamma^2/2 & \gamma & 0   \cr
\gamma & \gamma & 1  &  0 \cr
\gamma & \gamma  & 0 & 1 } .
\end{equation}
When applied to the momentum of a massless particle
moving in the negative $z$ direction with
\begin{equation}
(p, - p, 0, 0),
\end{equation}
it leaves the above four-momentum invariant, but on the other hand
performs a gauge transformation when applied to the four-potentials
of the electromagnetic field~(Kim et al. 1990).

\par

\section{Jones Vectors}\label{jones}

In studying polarized light propagating
along the $z$ direction, the traditional approach is to consider the $x$
and $y$ components of the electric fields.  Their amplitude ratio and
the phase difference determine the state of polarization.  Thus, we can
change the polarization either by adjusting the amplitudes, by changing
the relative phases, or both.  For convenience, we call the optical
device which changes amplitudes an ``attenuator'' and the device which
changes the relative phase a ``phase shifter.''
\par
The traditional language for this two-component light is the Jones-matrix
formalism which is discussed in standard optics
textbooks~(Hecht 1970).  In this formalism, the above two
components are combined into one column matrix with the exponential
form for the sinusoidal function
\begin{equation}\label{jvec11}
\pmatrix{E_{x} \cr E_{y}} =
\pmatrix{A \exp{\left\{i(kz - \omega t + \phi_{1})\right\}}  \cr
B \exp{\left\{i(kz - \omega t + \phi_{2})\right\}}} .
\end{equation}
This column matrix is called the Jones vector~(Jones 1941) .
\par
The Jones-matrix formalism starts with the projection operator~(Hecht 1970)
\begin{equation}\label{projec}
\pmatrix{1 & 0 \cr 0 & 0} ,
\end{equation}
applicable to the Jones vector of Eq.(\ref{jvec11}).  This operator
keeps the $x$ component and completely eliminates the $y$-component
of the electric field.
\par
This is an oversimplification of the real
world where the attenuation factor in the $y$ direction is greater
than that of the $x$ direction.  We shall replace this projection
operator by an attenuation matrix which is closer to the real world.
\par
In this note, we replace the projection operator of Eq.(\ref{projec})
by a squeeze matrix.
There are two transverse directions which are perpendicular to each
other.  The absorption coefficient in one transverse direction could
be different from the coefficient along the other direction.  Thus,
there is the ``polarization'' coordinate in which the absorption can
be described by~(Opatrny et al. 1993, Han et al. 1997, Ben-Aryeh 2005)
\begin{equation}\label{atten}
\pmatrix{e^{-\mu_{1}} & 0 \cr 0 & e^{-\mu_{2}}} =
e^{-(\mu_{1} + \mu_{2})/2} \pmatrix{e^{\mu/2} & 0 \cr 0 & e^{-\mu/2}}
\end{equation}
with $\mu = \mu_{2} - \mu_{1}$ .  Let us look at the projection
operator of Eq.(\ref{projec}).  Physically, it means that the absorption
coefficient along the $y$ direction is much larger than that of the $x$
direction.  The absorption matrix in Eq.(\ref{atten}) becomes the
projection matrix if $\mu_{1}$ is very close to zero and $\mu_{2}$
becomes infinitely large.  The projection operator of Eq.(\ref{projec})
is therefore a special case of the above attenuation matrix.

The attenuation matrix of Eq.(\ref{atten}) tells us that the electric
fields are attenuated at two different rates.  The exponential factor
$e^{-(\mu_{1} + \mu_{2})/2}$ reduces both components at the same rate
and does not affect the state of polarization.  The effect of
polarization is solely determined by the squeeze matrix
\begin{equation}\label{sq11}
B(\mu) = \pmatrix{e^{\mu/2} & 0 \cr 0 & e^{-\mu/2}} ,
\end{equation}
which is given in Eq.(\ref{core26}).  This type of mathematical operation
is quite familiar from studies of
squeezed states of light, if not from Lorentz boosts of spinors.  Thus,
we are expanding the Jones-matrix formalism by replacing the projection
operator of Eq.(\ref{projec}) by the squeeze operator in Eq.(\ref{sq11}).

\par

Another basic element is the optical filter with two different values
of the index of refraction along the two orthogonal directions.  The
effect on this filter can be written as
\begin{equation}\label{phase3}
\pmatrix{e^{i\delta_{1}} & 0 \cr 0 & e^{i\delta_{2}}}
= e^{i(\delta_{1} + \delta_{2})/2}
\pmatrix{e^{-i\delta/2} & 0 \cr 0 & e^{i\delta/2}} ,
\end{equation}
with $\delta = \delta_{1} - \delta_{2}$ .
In measurement processes, the overall phase factor
$e^{i(\delta_{1} + \delta_{2})/2}$
cannot be detected, and can therefore be deleted.  The polarization
effect of the filter is solely determined by the matrix
\begin{equation}\label{shif11}
Z(\delta) = \pmatrix{e^{i\delta/2} & 0 \cr 0 & e^{-i\delta/2}} ,
\end{equation}
which leads to a phase difference of $\delta$ between the $x$ and $y$
components.  The mathematical expression for this matrix is given in
Eq.(\ref{z22}).  It has a different physical meaning in the symmetry
of the Lorentz group.

\par

The polarization axes are not always the $x$ and $y$ axes.
For this reason, we need the rotation matrix
\begin{equation}\label{rot11}
R(\theta) = \pmatrix{\cos(\theta/2) & -\sin(\theta/2)
\cr \sin(\theta/2) & \cos(\theta/2)} .
\end{equation}
The traditional Jones-matrix formalism consists of systematic
combinations of the above three components given in Eq.(\ref{projec}),
Eq.(\ref{shif11}) and Eq.(\ref{rot11}).
\par

\subsection{Squeeze and Phase shift}

The effect of the phase shift matrix $Z(\delta)$ of Eq.(\ref{shif11}) on
the Jones vector is well known, but the effect of the squeeze matrix
of Eq.(\ref{sq11}) is not addressed adequately in the literature.  Let us
discuss the combined effect of these two matrices.  First of all both
are diagonal and they commute with each other.
\par
\begin{figure}
\centerline{\includegraphics[scale=0.24]{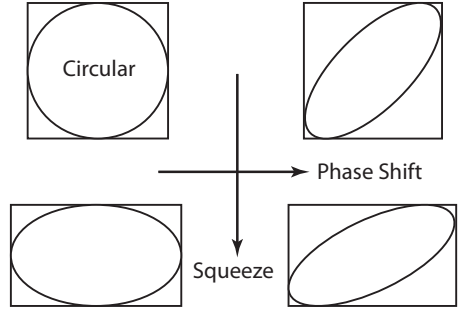}}
\vspace{2mm}
\caption{Squeeze and Phase Shift.  Both squeeze and phase shifts
result in elliptic deformations, but they are done differently.}\label{psq11}
\end{figure}

The effect of the squeeze matrix on the Jones vector is straight-forward
and the net result is
\begin{equation}\label{atonj}
\pmatrix{e^{\mu/2} & 0 \cr 0 & e^{-\mu/2}} \pmatrix{E_{x}  \cr E_{y}}
   = \pmatrix{e^{\mu/2} E_{x} \cr e^{-\mu/2} E_{y}} .
\end{equation}
This squeeze transformation expands one amplitude, while contracting the
other so that the product of the amplitudes remain invariant.   This
squeeze transformation is illustrated in Fig.~\ref{psq11}.

\par
In order to illustrate phase shifts, we start with the Jones vector
of the form
\begin{equation}
\pmatrix{\exp{(ikz)} \cr \exp{[i(kz - \pi/2)]} } ,
 \end{equation}
whose real part is
\begin{equation}
\pmatrix{x \cr y} = \pmatrix{\cos(kz) \cr \sin(kz)} ,
 \end{equation}
which corresponds to a circular polarization with
\begin{equation}
x^2 + y^2 = 1 .
\end{equation}
If we apply the phase shift matrix, the resulting vector is
\begin{equation}
\pmatrix{x \cr y} = \pmatrix{\cos(kz + \delta/2) \cr \sin(kz - \delta/2)} ,
 \end{equation}
which can be written as
\begin{equation}
\pmatrix{x \cr y} = \pmatrix{\cos(kz - \pi/4 + \alpha) \cr \cos(kz - \pi/4 -\alpha)} ,
 \end{equation}
with
\begin{equation}
\alpha = \frac{\delta}{2} + \frac{\pi}{4} .
\end{equation}
Then
\begin{eqnarray}
&{}& x + y = 2 (\cos\alpha) \cos(kz - \pi/4), \nonumber\\[1ex]
&{}& x - y = - 2 (\sin\alpha) \sin(kz - \pi/4),
\end{eqnarray}
and
\begin{equation}
\frac{(x + y)^2}{4(\cos\alpha)^2} + \frac{(x - y)^2}{4(\sin\alpha)^2} = 1 .
\end{equation}
This is an elliptic polarization.
\par
The squeeze operation of Eq.(\ref{sq11}) is  relatively simple.  It changes
the amplitudes, and it commutes with the phase shift matrix.  Thus, the
combined effect could be illustrated in Fig.~\ref{psq11}.

\subsection{Rotation of the Polarization Axes}\label{combine}

If the polarization coordinate is the same as the $xy$ coordinate where
the electric field components take the form of Eq.(\ref{jvec11}), the
attenuator is directly applicable to the Jones vector as in
Eq.(\ref{atonj}).  If the polarization coordinate is rotated by an angle
of $(\theta/2)$, or by the matrix
\begin{equation}
R(\theta) = \pmatrix{\cos(\theta/2) & -\sin(\theta/2)
\cr \sin(\theta/2) & \cos(\theta/2)} ,
\end{equation}
then the polarization coordinate makes an angle $(\theta/2)$ with
the $xy$ coordinate system. So the phase shifter takes the form
\begin{eqnarray}\label{pshif22}
&{}& Z(\theta, \delta) = R(\theta) P(\delta) R(-\theta) \\[1ex]
&{}& = \pmatrix{\cos(\delta/2) + i\sin(\delta/2)\cos\theta
    & i \sin(\delta/2) \sin\theta \cr    i \sin(\delta/2) \sin\theta
    &  \cos(\delta/2) - i\sin(\delta/2)\cos\theta } .
\end{eqnarray}
If the polarization coordinate system is rotated by $45^o$,
the phase shifter matrix becomes
\begin{equation}\label{pshif33}
Q(\delta) = \pmatrix{\cos(\delta/2)   & i\sin(\delta/2) \cr
  i\sin(\delta/2) & \cos(\delta/2) }
\end{equation}
\par

In order to illustrate what this matrix does to the polarized
beams, let us start with the circularly polarized wave
\begin{equation}
\pmatrix{ 1 \cr -i} e^{(ikz -i\omega t)} ,
\end{equation}
whose real part is
\begin{equation}
\pmatrix{ X \cr Y} = \pmatrix{ \cos(kz - \omega t) \cr \sin(kz - \omega t)}.
\end{equation}
This leads to the familiar equation for the circle
\begin{equation}
X^2 + Y^2 = 1 .
\end{equation}

\par
If the phase shifter of Eq.(\ref{pshif33}) is applied to the above
Jones vector, the result is
\begin{equation}
\pmatrix{ [\cos(\delta/2) + \sin(\delta/2)] \cos(kz- \omega t) \cr
 i[\sin(\delta/2) - \cos(\delta/2)] \sin(kz - \omega t)   }
\end{equation}
with
\begin{eqnarray}
&{}&\cos(\delta/2) = \cos\left([\delta/2 + \pi/4] - \pi/4\right), \nonumber\\[1ex]
&{}& \sin(\delta/2) = \cos\left([\delta/2 + \pi/4] + \pi/4\right) .
\end{eqnarray}
Thus,

\begin{eqnarray}
&{}&\cos(\delta/2) + \sin(\delta/2) =
              \sqrt{2}\cos\left(\delta/2 + \pi/4 \right) , \nonumber\\[1ex]
&{}& \cos(\delta/2) - \sin(\delta/2) = \sqrt{2} \sin\left(\delta/2 + \pi/4\right) .
\end{eqnarray}
After the phase shift, the Jones vector becomes
\begin{equation}
\pmatrix{ [\sqrt{2}\cos\alpha] \cos(kz- \omega t) \cr
 [\sqrt{2} \sin\alpha] \sin(kz - \omega t) },
\end{equation}
with
\begin{equation}
\alpha = \frac{\delta}{2} + \frac{\pi}{4} .
\end{equation}
The the $x$ and $y$ components will satisfy the equation
\begin{equation}
\frac{X^2}{(\sqrt{2} \cos\alpha)^2 } + \frac{Y^2}{(\sqrt{2} \sin\alpha)^2 } = 1.
\end{equation}
This is an elliptic polarization.

\begin{figure}
\centerline{\includegraphics[scale=0.24]{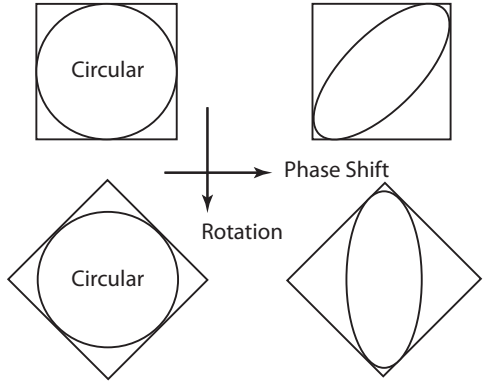}}
\vspace{2mm}
\caption{Phase shift and rotation. They are rotated by $45^o$.}\label{psq22}
\end{figure}

\par

Let us next consider rotations of the squeeze matrix
\begin{equation}\label{asq11}
B(\theta, \mu) = R(\theta) B(\mu) R(-\theta) ,
\end{equation}
which leads to
\begin{equation}\label{asq22}
B(\theta, \mu) = \pmatrix{\cosh(\mu/2) + \sinh(\mu/2) \cos\theta
& \sinh(\mu/2) \sin\theta \cr \sinh(\mu/2) \sin\theta
&\cosh(\mu/2) - \sinh(\mu/2) \cos\theta} .
\end{equation}
From Sec.~\ref{decomp}, we are familiar with this squeeze operation
which changes the amplitudes.

\par
If two squeeze transformations are made in two different directions,
the result is not another squeeze, but a squeeze matrix followed by
a rotation, which can be written as~(Ba\c{s}kal et al. 2005)
\begin{equation}\label{wrot11}
B(\theta,\lambda) B(0, \mu) = B(\phi, \xi) R(\omega),
\end{equation}
where
\begin{eqnarray}\label{tan11}
&{}& \cosh\xi = \cosh\mu~\cosh\lambda
            + \sinh\mu~\sinh\lambda~\cos\theta ,    \nonumber \\[2ex]
&{}&  \tan\phi = \frac{\sin\theta [\sinh\lambda +  \tanh\mu (\cosh\lambda -1) \cos\theta]
       }{\sinh\lambda \cos\theta + \tanh\mu [1 +
        (\cosh\lambda - 1)\cos^{2}\theta]} ,\nonumber \\[2ex]
&{}&  \tan\omega =  \frac{2 (\sin\theta) [\sinh\lambda\sinh\mu  +
  C_{-} \cos\theta] }{    C_{+} + C_{-}\cos(2\theta) +
     2 \sinh\lambda\sinh\mu\cos\theta } ,
\end{eqnarray}
with
\begin{equation}
C_{\pm} = (\cosh\lambda \pm 1) (\cosh\mu \pm 1) .
\end{equation}

\par

Indeed, Eq.(\ref{wrot11}) can be written as
\begin{equation}\label{wrot22}
R(\omega) = B(\phi, -\xi)~B(\theta, \lambda)~ B(0, \mu) ,
\end{equation}
three squeeze transformations lead to one rotation.
\par
We have
done this calculation using the kinematics of Lorentz
transformations.  On the other hand, it does not
appear possible to do experiments using high-energy particles.
However, it is gratifying to note that this experiment is
possible in polarization optics.
\par
If the angle $\theta$ is $90^o$
we use the notation $S(\lambda)$ for $B(\frac{\pi}{2},\lambda)$.  So
\begin{equation}\label{asq33}
S(\lambda) = \pmatrix{\cosh(\lambda/2)  & \sinh(\lambda/2) \cr
      \sinh(\lambda/2) & \cosh(\lambda/2) },
\end{equation}
and the above calculations become simpler with
\begin{equation}\label{wrot33}
S(\lambda) B(0,\mu) = B(\phi, \xi) R(\omega),
\end{equation}
where
\begin{eqnarray}\label{tan22}
&{}& \cosh\xi = \cosh\mu~\cosh\lambda ,    \nonumber \\[1ex]
&{}&  \tan\phi = \frac{\sinh\lambda}{\tanh\mu} ,\nonumber \\[1ex]
&{}&  \tan\omega = \frac{\sinh\lambda\sinh\mu}{\cosh\mu + \cosh\lambda} .
\end{eqnarray}

\subsection{Optical Activities}\label{opacti}

For convenience, let us change the parameters  $\theta$ and $\mu$ as
\begin{equation}
\theta = 2 \alpha z,  \qquad \mu = 2\beta z ,
\end{equation}
and the $R(\theta)$ and $S(\mu)$ matrices as
\begin{equation}\label{rot22}
R(\alpha z) = \pmatrix{\cos(\alpha z) & -\sin(\alpha z) \cr
  \sin(\alpha z) & \cos(\alpha z)} ,
\end{equation}
and the rotation angle increased as the beam propagates along the $z$ direction.
This version of optical activity is well known.
\par
In addition, we can consider the squeeze operation
\begin{equation}\label{sq22}
S(-\beta z) = \pmatrix{\cosh(\beta z) & -\sinh(\beta z) \cr
  -\sinh(\beta z) & \cosh(\beta z) } .
\end{equation}
Here the squeeze parameter increases as the beam moves.
The negative sign for $\beta$ is for convenience.

\par
If this squeeze is followed by the rotation of Eq.(\ref{rot22}),
the net effect is
\begin{equation}
\pmatrix{\cosh(\beta z) & -\sinh(\beta z) \cr
  -\sinh(\beta z) & \cosh(\beta z) }
\pmatrix{\cos(\alpha z) & -\sin(\alpha z) \cr
  \sin(\alpha z) & \cos(\alpha z)}
\end{equation}
where $z$ is in a macroscopic scale, perhaps measured at the order
of centimeters.  However, this is not an accurate description
of the optical process.

\par
In fact it happens in a microscopic scale of $z/N$,
and becomes accumulated into the macroscopic scale of $z$ after
$N$ repetitions, where $N$ is a very large number.  We are
thus led to the transformation matrix of the form~(Kim 2010)
\begin{equation}\label{trans}
M(\alpha,\beta,z)= \left[S(-\beta z/N)R(\alpha z/N)\right]^N .
\end{equation}
In the limit of large $N$, this quantity becomes
\begin{equation}
 \left[\pmatrix{1 & -\beta z/N \cr -\beta z/N & 1}
\pmatrix{1 & - \alpha z/N \cr \alpha z/N & 1}\right]^N .
\end{equation}
Since $\alpha z/N$ and $\beta z/N$ are very small,
\begin{equation}\label{mat11}
M(\alpha,\beta,z)=  \left[\pmatrix{1 & 0 \cr 0 & 1}
   + \pmatrix{0 & -(\alpha + \beta) \cr
  (\alpha - \beta) & 0}\frac{z}{N} \right]^N .
\end{equation}
For large $N$, we can write this matrix as
\begin{equation}\label{m33}
M(\alpha,\beta,z) =  \exp{\left( H z \right)} ,
\end{equation}
with
\begin{equation}\label{g11}
H = \pmatrix{0 & -(\alpha + \beta) \cr (\alpha - \beta) & 0} .
\end{equation}

We can compute this matrix using the procedure developed in
Sec.~\ref{decomp}. If $\alpha$ is greater than $\beta$, $H$
becomes
\begin{equation}\label{g22}
  H = \alpha' \pmatrix{0 & \exp{(\eta)} \cr \exp{(-\eta)}  & 0} ,
\end{equation}
with
\begin{eqnarray}\label{alpha}
&{}&  \alpha' = \sqrt{\alpha^2 - \beta^2}, \nonumber \\[1ex]
&{}&  \exp{(\eta)} = \sqrt{\frac{\alpha + \beta}{\alpha - \beta}},
\end{eqnarray}
and the $M$ matrix of Eq.(\ref{m33}) take the form
\begin{equation}
\pmatrix{\cos(\alpha' z)  &  -e^{\eta}\sin(\alpha' z) \cr
 e^{-\eta}\sin(\alpha' z) & \cos(\alpha' z)}
\end{equation}

If $\beta$ is greater than $\alpha$, the off-diagonal elements have
the same sign.  We can then write $H$ as
\begin{equation}\label{g33}
 H = - \beta' \pmatrix{0 & \exp{(\eta)} \cr \exp{(-\eta)}  & 0} ,
\end{equation}
with
\begin{eqnarray}\label{muk}
&{}&  \beta' =  \sqrt{\beta^2 - \alpha^2}, \nonumber\\[1ex]
&{}& \exp{(\eta)} = \sqrt{\frac{\beta + \alpha}{\beta - \alpha}} ,
\end{eqnarray}
and the $M$ matrix of Eq.(\ref{m33}) becomes
\begin{equation}
\pmatrix{\cosh(\beta' z)  &  -e^{\eta}\sinh(\beta' z) \cr
 -e^{-\eta}\sinh(\beta' z) & \cosh(\beta' z)} .
\end{equation}

\par

If $\alpha = \beta$, the lower-left element of the $H$ matrix has
to vanish, and it becomes
\begin{equation}
H = \pmatrix{0 & - 2\alpha \cr 0 & 0} ,
\end{equation}
and the $M$ matrix takes the triangular form
\begin{equation}
    \pmatrix{ 1 & -2\alpha z \cr 0 & 1}.
\end{equation}
\par
The optical material can be made to provide rotations of the
polarization axis.  It is much more interesting to see this
additional effect of squeeze.

\section{Stokes Parameters and the Poincar\'e Sphere}\label{stokes}

In Sec.~\ref{jones}, we studied various aspects of the Jones vector
formalism of the polarized beams, we have not dealt with the problem
of whether the two beams are coherent with each other.  In order
to study this coherence problem we have to construct the four
Stokes parameters.
\par
Let us write the Jones vector of Eq.(\ref{jvec11}) as
\begin{equation}\label{jvec22}
\pmatrix{\psi_{1} \cr \psi_{2}} =
\pmatrix{a~ \exp{\left\{i(kz - \omega t + \delta_1)\right\}} \cr
    b~\exp{\left\{i(kz - \omega t + \delta_2)\right\}} } ,
\end{equation}
where $a$ and $b$ are positive real numbers.
In Sec.~\ref{jones}, we studied the effects of the squeeze $B(\eta)$,
phase shift $Z(\delta)$, and rotation $R(\theta)$ on the Jones matrix.
These matrices are given in Eq.(\ref{sq11}), Eq.(\ref{shif11}) and
Eq.(\ref{rot11}) respectively.

\par
These transformation matrices can be written as one expression as the
two-by-two matrix of $G$ of Eq.(\ref{alphabeta}), and its role in the
Lorentz group and its physical application to the Jones vectors were
discussed in Secs.~\ref{sptime} and \ref{jones}, respectively.
While the Jones vector can deal with two independent beams, it does not
address the issue of whether they are coherent with other.  For this
purpose, let us introduce the coherency matrix~(Brosseau 1998, Saleh et al. 2007).
\begin{equation}\label{cocy11}
C = \pmatrix{S_{11} & S_{12} \cr S_{21} & S_{22}},
\end{equation}
with
\begin{equation}
<\psi_{i}^* \psi_{j}> = \frac{1}{T} \int_{0}^{T}\psi_{i}^* (t + \tau) \psi_{j}(t) dt,
\end{equation}
where $T$ is for a sufficiently long time interval, is much larger than $\tau$.
Then, those four elements become
\begin{eqnarray}
&{}& S_{11} = <\psi_{1}^{*}\psi_{1}> =a^2  , \qquad
S_{12} = <\psi_{1}^{*}\psi_{2}> = ab~e^{-(\sigma +i\delta)} , \nonumber \\[1ex]
&{}& S_{21} = <\psi_{2}^{*}\psi_{1}> = ab~e^{-(\sigma -i\delta)} ,  \qquad
S_{22} = <\psi_{2}^{*}\psi_{2}>  = b^2 .
\end{eqnarray}
The diagonal elements are the absolute values of $\psi_1$ and $\psi_2$
respectively.  The off-diagonal elements could be smaller than the
product of $\psi_1$ and $\psi_2$, if the two beams are not completely
coherent.  Thus, the parameter $\sigma$ serves as the decoherence
parameter.

\par
The $\sigma$ parameter specifies the degree of coherency.
Unlike the $ABCD$ matrix, this coherency matrix is not always real,
and its determinant is not always one.  If we restrict the trace of
this matrix to be one by normalizing, this matrix becomes the
density matrix~(Feynman 1972).
\par

If we start with the Jones vector of the form of Eq.(\ref{jvec22}),
the coherency matrix becomes
\begin{equation}\label{cocy22}
C = \pmatrix{a^2 & ab~e^{-(\sigma + i\delta)} \cr
ab~e^{-(\sigma - i\delta)} & b^2} .
\end{equation}
We are interested in the symmetry properties of this matrix.  Since
the transformation matrix applicable to the Jones vector is the
two-by-two representation of the Lorentz group, we are
particularly interested in the transformation matrices applicable to
this coherency matrix.
\par
The trace and the determinant of the above coherency matrix
are
\begin{eqnarray}
&{}& det(C) = (ab)^2 \left(1 - e^{-2\sigma}\right), \nonumber \\[2ex]
&{}& tr(C) = a^2 + b^2 .
\end{eqnarray}
Since $e^{-\sigma}$ is always smaller than one, we can introduce
an angle $\chi$ defined as
\begin{equation}
\cos\chi = e^{-\sigma} ,
\end{equation}
and call it the ``decoherrence angle.''  If $\chi = 0$, the decoherence is minimum,
and it is maximum when $\chi = 90^o$.  We can then write the decoherency
matrix of of Eq.(\ref{cocy22}) as
\begin{equation}\label{cocy22b}
C = \pmatrix{a^2 & ab(\cos\chi)e^{-i\delta} \cr
ab(\cos\chi)e^{i\delta} & b^2}.
\end{equation}

\par
The degree of polarization is defined as~(Saleh et al. 2007)
\begin{equation}
P = \sqrt{ 1 - \frac{4~det(C)}{(tr(C))^2}} =
        \sqrt{1 - \frac{4(ab\sin\chi)^2)}{(a^2 + b^2)^2}} .
\end{equation}
This degree is one if $\chi = 0$.  It becomes
\begin{equation}
  \frac{a^2 - b^2}{a^2 + b^2} ,
\end{equation}
when $\chi=90^o$.  We can without loss of generality assume that
$a$ is greater than $b$ . If they are equal, the degree of polarization
is zero.
\par

\subsection{Stokes Parameters as Four-Vectors}

Under the influence of the $G$ transformation given in Eq.(\ref{alphabeta}),
this coherency matrix is transformed as
\begin{eqnarray}\label{trans22}
&{}& C' = G~C~G^{\dagger} =
\pmatrix{S'_{11} & S'_{12} \cr S'_{21} & S'_{22}} \nonumber \\[2ex]
&{}&\hspace{5ex} = \pmatrix{\alpha & \beta \cr \gamma & \delta}
\pmatrix{S_{11} & S_{12} \cr S_{21} & S_{22}}
\pmatrix{\alpha^{*} & \gamma^{*} \cr \beta^{*} & \delta^{*}} .
\end{eqnarray}
Here, the $G$ matrix is not Hermitian, and its Hermitian conjugate is
not always its inverse.  Thus it is not a similarity transformation, yet
it preserves the determinant of $C$.
When the $ G $ matrix in Eq.(\ref{alphabeta})
consists of real elements, it becomes the transformation matrix applicable to the
$ABCD$ matrix.  If it is constrained to be Hermitian, it becomes a rotation
matrix without boosts.

\par

While the coherency matrix is transformed as in Eq.(\ref{trans22}),
its components transforms in the same manner as in Eq.(\ref{translate})
\begin{equation}\label{trans44}
\pmatrix{S_{11}' \cr S_{12}' \cr S_{21}' \cr S_{22}'} =
\pmatrix{\alpha\alpha^{*} & \alpha \beta^{*} &
\beta\alpha^{*} & \beta \beta^{*} \cr
\alpha \gamma^{*} & \alpha \delta^{*} &
\beta \gamma^{*} & \beta \delta^{*} \cr
\gamma \alpha^{*}  & \gamma \beta^{*} &
\delta \alpha^{*} & \delta \beta^{*} \cr
\gamma \gamma^{*} & \gamma \delta^{*} &
\delta \gamma^{*} & \delta \delta^{*}}
\pmatrix{S_{11} \cr S_{12} \cr S_{21} \cr S_{22}}  .
\end{equation}

\par
Particular combinations of the coherency matrix components of
Eq.(\ref{cocy11}) are crucial for the quantum picture of
polarization~(Falkoff et al. 1951):
\begin{eqnarray}\label{stokes11}
&{}& S_{0} = \frac{S_{11} + S_{22}}{\sqrt{2}},  \qquad
    S_{3} = \frac{S_{11} - S_{22}}{\sqrt{2}},    \nonumber \\[2ex]
&{}& S_{1} = \frac{S_{12} + S_{21}}{\sqrt{2}}, \qquad
S_{2} = \frac{S_{12} - S_{21}}{\sqrt{2} i}
\end{eqnarray}
which can also be expressed as
the sum of an identity matrix and the Pauli spin matrices
$\sigma^{i}$~(Fano 1954)
\begin{equation}
C=\frac{1}{2}(S_{0}I+S_{1}\sigma^{1}+S_{2}\sigma^{2}+S_{3}\sigma^{3})
\end{equation}
where the coefficients are known as the Stokes parameters in the
literature~(Shurcliff 1962).  Furthermore, expressing in such a compact
form also serves to treat the Jones and Mueller calculi in the
framework of pure operational Pauli algebraic approach~(Tudor 2010).

\par
The the four-by-four matrix which transforms
$\left(S_{11}, S_{22}, S_{12}, S_{21} \right)$ to
$\left(S_{0}, S_{3}, S_{1}, S_{2} \right)$ is
\begin{equation}
\pmatrix{S_{0} \cr S_{3} \cr S_{1} \cr  S_{2} } = \frac{1}{\sqrt{2}}
 \pmatrix{1 & 1 & 0 & 0 \cr 1 & -1 & 0 & 0 \cr
 0 & 0 & 1 & 1 \cr  0 & 0 & -i & i}
\pmatrix{S_{11} \cr S_{22} \cr S_{12} \cr  S_{21} } .
\end{equation}
This matrix enables us to construct the transformation matrix
applicable to the Stokes parameters, widely known as the Mueller
matrix~(Soleillet 1929, Mueller 1943, Brosseau 1998).
\par

The Mueller matrix applicable to the Stokes parameters takes the
same form as the Lorentz transformation matrix applicable to the
space-time four-vector of $(t, z, x, y)$ given in Sec.~\ref{sptime}.
Therefore, the Mueller matrix is a four-by-four representation of
the Lorentz group.

\par
It is gratifying to note that the four-by-four Mueller matrices
share the same symmetry properties as those of the two-by-two
Jones matrices applicable to the Jones vectors.
Thanks to the squeeze matrix $B(\mu)$ of Eq.(\ref{sq11}), we are
able to extend the symmetry of those two-by-two matrices from $SU(2)$
to $SL(2,c)$~(Han et al. 1997, Devlaminck et al. 2008, Redkov 2011).

\par
We should note here that the decoherence angle is a Lorentz-invariant
quantity.  It cannot be changed by Mueller transformations.  It may
be possible to construct a four-by-four matrix which will change this
parameter~(Ortega-Qujiano et al. 2011), but this matrix cannot belong to
the Lorentz group.
\par
As for the two-by-two matrix formalism, it is an iterating proposition
to formulate the problem using
quarternions~(Dlugunovich et al. 2009, Tudor 2010).
Quarternions represent a four-dimensional rotation group and do
more.  Thus, interesting results may be obtained from this line
of approach.

\subsection{Winger's Little Group for Internal Space-time Symmetries}
It is more interesting to study the problem using the two-by-two
representation of the coherency matrix because their
elements are directly measurable quantities.  For this purpose,
let us recall the Lorentz transformation of the four-vector of a free
particle $\left(p_0, p_3, p_1, p_2 \right)$ which is the same as that
of the $(t, z, x, y)$ four-vector, and we can write its $G$
transformation as
\begin{eqnarray}\label{trans33}
&{}& P' = G~P~G^{\dagger} =
\pmatrix{p'_0 + p'_z & p'_x - ip'_y \cr p'_x + ip'_y &  p'_0 - p'_z} \nonumber \\[2ex]
&{}&\hspace{5ex} = \pmatrix{\alpha & \beta \cr \gamma & \delta}
\pmatrix{p_0 + p_z &  p_x - ip_y \cr p_x + ip_y &  p_0 - p_z}
\pmatrix{\alpha^{*} & \gamma^{*} \cr \beta^{*} & \delta^{*}} .
\end{eqnarray}
The two-by-two matrix for the four-momentum $P$ is given in Eq.(\ref{slc05}).

\par
We can consider transformations which will leave the four-momentum invariant.
In other words, we can write Wigner's little group as the subset of the $G$
matrix which satisfies
\begin{equation}\label{trans55}
\pmatrix{p_0 + p_z & p_x - ip_y \cr p_x + ip_y &  p_0 - p_z}
 = \pmatrix{\alpha & \beta \cr \gamma & \delta}
\pmatrix{p_0 + p_z &  p_x - ip_y \cr p_x + ip_y &  p_0 - p_z}
\pmatrix{\alpha^{*} & \gamma^{*} \cr \beta^{*} & \delta^{*}} .
\end{equation}
Using the rotation matrix $Z(\delta)$ of Eq.(\ref{shif11}) which leads to
a phase shift, we can bring this formula to the form
\begin{equation}\label{trans55b}
\pmatrix{p_0 + p_z & p_x \cr p_x  &  p_0 - p_z}
 = \pmatrix{\alpha & \beta \cr \gamma & \delta}
\pmatrix{p_0 + p_z &  p_x  \cr p_x  &  p_0 - p_z}
\pmatrix{\alpha & \gamma \cr \beta & \delta} .
\end{equation}
This can then be transformed to a diagonal form
\begin{equation}\label{trans66}
\pmatrix{p_0 + \sqrt{p_z^2 + p_x^2} & 0 \cr 0 &  p_0 - \sqrt{p_z^2 + p_x^2}},
\end{equation}
with the rotation matrix $R(\xi)$, where
\begin{equation}
    \tan\xi = \frac{p_x}{p_z} .
\end{equation}
\par
With the boost squeeze matrix $B(\eta)$ given in Eq.(\ref{sq11}) where
\begin{equation}
e^{\eta} = \sqrt{\frac{p_0 + p_z}{p_0 - p_z}} ,
\end{equation}
we can transform the diagonal matrix of Eq.(\ref{trans66}) to another diagonal
matrix
\begin{equation}
\pmatrix{m & 0 \cr 0 & m},
\end{equation}
where $m = \sqrt{p_o^2 - p_z^2 - p_x^2}.$  The squeeze matrix $B(\eta)$
corresponds the Lorentz boost given in Eq.(\ref{4by4}).

\par
The $G$ transformation matrix which will leave this four-momentum matrix is
has to be Hermitian, and is a rotation matrix $R(\theta)$.  This defines
Wigner's internal space-time symmetry of a massive particle.
\par
This matrix remains invariant under the $G$ transformation if the $G$
matrix is Hermitian.  It is Hermitian only for rotations.  This is thus
consistent with Wigner's $O(3)$-like little group for massive particles,
as discussed in Sec.~\ref{sptime}.
\par
For a massless particle, we can choose the system where $p_0 = p_z = \omega$
and $p_x = p_y = 0$.  Then the $P$ matrix becomes
\begin{equation}\label{wlg22}
P = \pmatrix{2\omega & 0 \cr 0 & 0} ,
\end{equation}
and its determinant is zero, saying $p_0^2 - p_z^2 = 0$.  It is not
difficult to construct the $G$ matrix whose $G$ transformation will leave
the above $P$ matrix invariant.  It takes triangular form
\begin{equation}\label{gamma22}
 \pmatrix{1 & \gamma \cr 0 & 1} .
\end{equation}
From this, it is not difficult to construct its four-by-four counterpart
given in Eq.(\ref{gamma44}).
\par

Let us go back to the symmetry of the coherency matrix, and to the
matrix $C$ of Eq.(\ref{cocy22}).  If we make a $G$ transformation with
$Z(\delta)$, the $D$ matrix becomes
\begin{equation}
\pmatrix{a^2 &  ab~\cos\chi \cr ab~\cos\chi & b^2} ,
\end{equation}
The two-by-two matrix of $Z(\delta)$ is given in Eq.(\ref{shif11}).
\par
If we make another $G$ transformation with $R(-\theta)$ where
\begin{equation}
\tan\theta  = \frac{2ab~\cos\chi}{a^2 - b^2},
\end{equation}
the coherency matrix becomes
\begin{equation}\label{cocy33}
C = \pmatrix{s + r  & 0 \cr 0 & s - r} ,
\end{equation}
with
\begin{eqnarray}
&{}& s = \frac{1}{2} \left(a^2 + b^2 \right) , \nonumber \\[1ex]
&{}& r = \frac{1}{2} \sqrt{\left(a^2 + b^2\right)^2 + 4(ab)^2 \sin^2\chi}.
\end{eqnarray}

 \par
If $\theta = 0$, the system is totally coherent, and the coherency matrix
becomes
\begin{equation}\label{cocy44}
\pmatrix{a^2 + b^2 & 0 \cr 0 & 0 } ,
\end{equation}
This matrix is like the four-momentum matrix of Eq.(\ref{slc05}) for
massless particles.

\par
If $\chi$ is nonzero, we can $G$-transform the $D$ matrix of with
\begin{equation}
B(-\eta) = \pmatrix{e^{-\eta/2} & 0 \cr 0 & e^{\eta/2} } ,
\end{equation}
with
$$
e^\eta = \sqrt{\frac{s + r}{s - r}} ,
$$
the $D$ matrix becomes
\begin{equation}
C = \pmatrix{\sqrt{s^2 - r^2} & 0 \cr 0 & \sqrt{s^2 - r^2} } ,
\end{equation}
This coherency matrix is invariant under $G$ transformations if the $G$
matrix consists only of rotations and thus is Hermitian.  This aspect
is consistent with Wigner's $O(3)$-like little group for massive
particles.
\par

In the case of the four-momentum matrix, its determinant is $m^2$ and
is Lorentz-invariant.  For the coherency matrix, the determinant is
$(ab)^2\sin^2\chi$.  This means that the coherency angle $\chi$ cannot be
changed by Lorentz transformations, as in the case of mass in special
relativity.

\section{Symmetries of the Poincar\'e Sphere}\label{poins}

The Poincar\'e sphere has a long history, and its spherical symmetry is
well known~(Born et al. 1980).  The rotational symmetry of the
Poincar\'e sphere is readily included in the Lorentz symmetry.
We shall first review the rotational symmetry, and study the
effect of Lorentz boosts.
\par

Let us write the coherency matrix of Eq.(\ref{cocy22}) as a
four-component vector
\begin{equation}\label{4vec33}
\pmatrix{s \cr r_z \cr r_x \cr r_y}
= \pmatrix{s \cr r(\cos\theta) \cr
        r~(\sin\theta)\cos\delta \cr r(\sin\theta)\sin\delta}
=  \pmatrix{ (a^2 + b^2)/2  \cr (a^2 - b^2)/2 \cr
ab(\cos\delta)\cos\chi \cr ab(\sin\delta)\cos\chi } .
\end{equation}
This four-vector is defined by two spheres: the outer sphere with radius $s$,
where
\begin{equation} \label{radius11}
s =\frac{(a^2 + b^2)}{2} ,
\end{equation}
and the inner sphere defined by the three-component vector
$\left(r_z,~r_x,~r_y \right)$, with its radius
\begin{eqnarray}\label{radius22}
&{}& r = \sqrt{r_z^2 + r_x^2 + r_y^2}  \nonumber \\[2ex]
&{}& \hspace{3mm} =\frac{1}{2}
                   \sqrt{\left(a^2 - b^2\right)^2 + 4 (ab)^2 \cos^{2}\chi }
\end{eqnarray}
which is the radius of the Poincar\'{e} sphere.  Its $z$ component is
\begin{equation}
r_z = \frac{a^2 - b^2}{2} ,
\end{equation}
which is independent of the decoherency angle $\chi$.  Here we
assume the amplitude $a$ to be greater than $b$.

\par
\begin{figure}[thb]
\centerline{\includegraphics[scale=0.20]{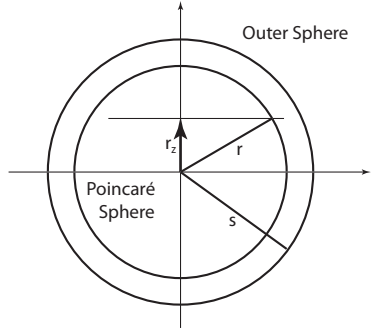}}
\vspace{2mm}
\caption{Poincar\'e sphere and its outer sphere.  The radius of the
 Poincar\'{e} sphere depends on the decoherence angle $\chi$, but
 its z component does
 not. }
\label{poinc11}
\end{figure}

The radius of the Poincar\'e sphere $r$ depends on $\chi$, and takes its
maximum value $s$  when $\chi= 0$.  The radius shrinks to its minimum
value $r_z$  when $\chi$ goes to it largest value.  Figure~\ref{poinc11}
illustrates the circles and their radii.  The ratio of $r_z$ to $r$ is
\begin{equation}\label{ratio11}
\cos\theta = \frac{r_z}{r} = \frac{a^2 -b^2}
            {\sqrt{ (a^2 - b^2)^2 +4(ab)^2 \cos^{2}\chi}  }.
\end{equation}

\par

If we apply the rotation
\begin{equation}
\pmatrix{1 & 0 & 0 &  0 \cr
         0 & 1 & 0 &  0 \cr
         0 & 0 & \cos\delta &  \sin\delta \cr
         0 & 0 & -\sin\delta &  \cos\delta } ,
\end{equation}
and then
\begin{equation}
\pmatrix{1 & 0 & 0 &  0 \cr
         0 & \cos\theta &  \sin\theta & 0 \cr
          0 & -\sin\theta &  \cos\theta & 0 \cr
         0 & 0 & 0 &  1 } .
\end{equation}
with $\cos\theta = r_z / r $ given in Eq.(\ref{ratio11}), this four-vector
can be brought into the form
\begin{equation}\label{rps1}
(s,~r,~0,~0).
\end{equation}

\par

Within the framework of the traditional three-dimensional
geometry of the Poincar\'e sphere, it is possible to transform the
four vector of Eq.(\ref{4vec33}) to the four vector of Eq.(\ref{rps1})
while the value of $s$ is left unchanged.  On the other hand, the Lorentz
symmetry allows a transformation on this four-vector by $B(-\eta)$
with $\tanh\eta = r/s$ so that the four-vector
becomes~(Ba\c{s}kal et al. 2006)
\begin{equation}\label{lt66}
\pmatrix{\sqrt{s^2 - r^2} \cr 0 \cr 0 \cr 0} =
  \pmatrix{\cosh\eta & -\sinh\eta & 0 & 0 \cr
      -\sinh\eta & \cosh\eta & 0 & 0 \cr
       0 & 0 & 1 & 0 \cr  0 & 0 & 0 & 1 }
  \pmatrix{s \cr r \cr 0 \cr 0}.
\end{equation}
This means that the radius of the Poincar\'e sphere can become
zero while the outer radius takes its minimum value
$ab~\sin\chi$.

\par

Indeed, the angle $\chi$ determines the minimum radius of the
outer sphere.  This radius takes the maximum value of $ab$
when $\sigma$ becomes infinity.  The larger radius becomes zero
when $\sigma = 0$ corresponding to completely coherent beams,
and it does not correspond to the real world.
\par
In order to resolve this problem, we note that the large radius
and smaller radius are the same and its value is
$\left(a^2 + b^2\right)/2$, when $\chi = 0$.  Thus, if we
make the Lorentz boost of Eq.(\ref{lt66}), the result is
\par
\begin{equation}\label{lt77}
e^{-\eta}~\pmatrix{s \cr s \cr 0 \cr 0} =
  \pmatrix{\cosh\eta & -\sinh\eta & 0 & 0 \cr
      -\sinh\eta & \cosh\eta & 0 & 0 \cr
       0 & 0 & 1 & 0 \cr  0 & 0 & 0 & 1 }
  \pmatrix{s \cr s \cr 0 \cr 0}.
\end{equation}

The Lorentz group, including its rotation subgroup, changes all
the parameters for the coherency matrix.  However, it cannot change
the decoherence angle $\chi$.  What significance does this carry from
the symmetry point of view and from the physical point of view?

\subsection{O(3,2) Symmetry of the Poincar\'e Sphere}\label{o32}
It is clear from the previous subsection that the decoherence angle
$\chi$ is invariant under Lorentz transformations.  We are now looking
for another symmetry which will change this variable.  For this purpose
we write the coherency matrix of Eq.(\ref{cocy22b})
as
\begin{equation}\label{cocy55}
C_1 = \pmatrix{a^2 & ab(\cos\chi)e^{-i\delta} \cr
ab (\cos\chi)e^{i\delta} & b^2}
\end{equation}
and introduce another matrix where $\cos\chi$ is replaced by $\sin\chi$,
which can take the form
\begin{equation}\label{cocy66}
C_2 = \pmatrix{a^2 & ab(\sin\chi)e^{-i\delta} \cr
ab (\sin\chi)e^{i\delta} & b^2} .
\end{equation}

\par
For the first coherency matrix of Eq.(\ref{cocy55}), we know how to
transform it into the four-vector $(s,~r_1,~0,~0),$  with
\begin{equation}
 s = \frac{a^2 + b^2}{2}, \qquad
    r_1 = \frac{1}{2}\sqrt{(a^2 - b^2)^2 - 4(ab)^2(\cos\chi)^2 } .
\end{equation}
Then, the second matrix can be brought to the four-vector
$(u,~r_2,~0,~0),$ with
\begin{equation}
 u = \frac{a^2 + b^2}{2}, \qquad
    r_2 = \frac{1}{2}\sqrt{(a^2 - b^2)^2 - 4(ab)^2(\sin\chi)^2 } .
\end{equation}

\par

These two expressions lead us to consider the following two four
vectors.
\begin{equation}\label{4vec55}
\pmatrix{ab(\cos\chi) \cr 0 \cr 0 \cr 0 } , \qquad
\pmatrix{ab(\sin\chi) \cr 0 \cr 0 \cr 0 } .
\end{equation}
In view of the relation $ \cos^2\chi + \sin^2\chi = 1$, we are led
to the five dimensional vector space with
$\left(s,~u,~r_z,~r_x,~r_y \right)$, which can start with
\begin{equation}
\left(ab\cos\chi,~ab\sin\chi,~0,~0,~0\right).
\end{equation}
Now we can change the value of the decoherence parameter $\sigma$
by changing the angle $\chi$, but we can change this variable by
introducing a rotation matrix applicable to the two-dimensional
vector space of $s$ and $t$.  When  all other components vanish
we can write the rotation matrix
\begin{equation}
  \pmatrix{ab(\cos\chi) \cr ab(\sin\chi)}
    = \pmatrix{\cos\chi & -\sin\chi \cr \sin\chi & \cos\chi}
   \pmatrix{ab \cr 0} .
\end{equation}
Therefore the five-by-five rotation matrix will be of the form
\begin{equation}\label{tu}
\pmatrix{\cos\chi & -\sin\chi & 0 & 0 & 0  \cr
        \sin\chi & \cos\chi & 0 & 0 & 0 \cr
          0 & 0 & 1 & 0 & 0 \cr
          0 & 0 & 0 & 1 & 0 \cr
          0 & 0 & 0 & 0 & 1},
\end{equation}
which is applicable to the five-component vector
$\left(s,~t,~r_z,~r_x,~r_y \right)$,
with the two four-dimensional subspaces, corresponding to the coherency
matrices
\begin{eqnarray}\label{cocy77}
&{}& C_s(\chi) = \pmatrix{a^2 & ab\,e^{-i\delta}(\cos\chi)
\cr  ba\,e^{i\delta}(\cos\chi) &  b^2} ,      \nonumber \\[2ex]
&{}& C_t(\chi) = \pmatrix{a^2 & ab~e^{-i\delta}(\sin\chi)
\cr  ba~e^{i\delta}(\sin\chi) &  b^2} .
\end{eqnarray}
Here both $s^2 - r^2$ and $t^2 - r^2$ remain invariant under Lorentz
transformations within their four-dimensional subspaces.  Their
deteminants are
\begin{equation}\label{cocy88}
det(C_s) = (ab)^2~\sin^2\chi, \qquad
det(C_u) = (ab)^2~\cos^2\chi,
\end{equation}
resuting in
\begin{equation}\label{cocy99}
det\left(C_s\right) + det\left(C_s\right) = (ab)^2,
\end{equation}
which is independent of the angle $\chi$.

\subsection{Entropy Probem}\label{entro}
Another important way to measure the lack of coherence is to calculate
the entropy of the system.  The coherency matrices defined in this
section become density matrices if their traces are normalized
to be one.  Thus from Eq.(\ref{cocy66}), we can write
\begin{eqnarray}\label{den11}
&{}& \rho_s(\chi) = \frac{1}{a^2 + b^2}   \pmatrix{a^2 & ab\,e^{-i\delta}(\cos\chi)
\cr  ba\,e^{i\delta}(\cos\chi) &  b^2} ,      \nonumber \\[2ex]
&{}& \rho_u(\chi) = \frac{1}{a^2 + b^2} \pmatrix{a^2 & ab~e^{-i\delta}(\sin\chi)
\cr  ba~e^{i\delta}(\sin\chi) &  b^2} .
\end{eqnarray}
These matrices can be diagonalized to
\begin{eqnarray}\label{den22}
&{}& \rho_s(\chi) = \frac{1}{2} \pmatrix{1 + f_s  & 0  \cr  0 &  1 - f_s} ,      \nonumber \\[2ex]
&{}& \rho_u(\chi) = \frac{1}{2} \pmatrix{1 + f_u  & 0  \cr  0 &  1 - f_u} ,
\end{eqnarray}
where
\begin{equation}
f_s = \sqrt{1 - \frac{4(ab)^2 \sin^2\chi}{a^2 + b^2}}, \qquad
f_u = \sqrt{1 - \frac{4(ab)^2 \cos^2\chi}{a^2 + b^2}},
\end{equation}
\par
Then, their entropies become
\begin{eqnarray}
&{}& S_s = -\left(\frac{1 + f_s}{2} \right) \ln{\left(\frac{1 + f_s}{2}\right)}
 - \left(\frac{1 - f_s}{2} \right) \ln{\left(\frac{1 - f_s}{2}\right)}, \nonumber\\[2ex]
&{}& S_u = -\left(\frac{1 + f_u}{2} \right) \ln{\left(\frac{1 + f_u}{2}\right)}
         - \left(\frac{1 - f_u}{2} \right) \ln{\left(\frac{1 - f_u}{2}\right)} .
\end{eqnarray}
The entropy $S_s$ becomes zero when $\chi = 0$. It becomes
\begin{equation}
 \frac{a^2}{a^2 + b^2} \ln{\left(\frac{a^2 + b^2}{a^2}\right)}
 + \frac{b^2}{a^2 + b^2} \ln{\left(\frac{a^2 + b^2}{b^2}\right)},
\end{equation}
when $\chi = 90^o$.
The entropy $S_s$ is a monotonically increasing function of the angle
$\chi$ starting from zero to the above maximum value, which becomes $\ln{2}$
when $a = b$.

The entropy $S_{s}$ of the first space is monotonically increasing function
of $\chi,$ while that of the second space $S_{u}$ is a decreasing function.
Thus, an increase in entropy in the first space leads to a decrease in
the second space.  Then we can ask whether the sum of these two entropies
becomes independent of $\chi,$ leading to an entropy conservation of the
total system.  The answer is No.  However, this does not cause problems
for us, because the second space is not necessarily a physical space.
It could be meaningless to use the same definition of entropy for this
space.  On the other hand, as we noted before,  we can define the
conservation of entropy in terms of the the sum of the
determinants of the coherency matrices given in
Eq.(\ref{cocy77}).  Furthermore, this determinant condition
does not require that the amplitudes of the two beams to be the same.

\subsection{Feynman's Rest of the Universe}

What is the meaning of this second space?  In his book on statistical
mechanics, Feynman makes the following statement about
the density matrix~(Feynman 1972):
{\it When we solve a quantum-mechanical problem,
what we really do is divide the universe into two parts - the system in
which we are interested and the rest of the universe.  We then usually
act as if the system in which we are interested comprised the entire
universe.  To motivate the use of density matrices, let us see what
happens when we include the part of the universe outside the system}.

Feynman did not specify whether the rest of the universe is observable
or not.  In either case, it is an interesting exercise to construct
a model of the rest of the universe behaving like a physical world.
With this point in mind, one of us studied two coupled harmonic
oscillators in which one of the oscillators correspond to the physical
world and the other to the rest of the universe~(Han et al. 1999).
In this example, the rest of the universe is the same as the world
in which we do physics.  In thermal field theory~(Umezawa et al. 1982),
even though based on the same mathematics as that of the coupled
oscillators, the rest of the universe is not physically identified,
except that it causes thermal excitations of the oscillators in the
physical world.

The concept of decoherence occupies one of the central places
in the current development of physics. In~(Feynman et al. 1963)
when the system couples to finite temperature baths the result
is an environmentally induced decoherence.
Its effects can be determined in tunneling processes~(Caldeira et al. 1983)
and in two-state systems that are coupled to dissipative
environments~(Leggett et al. 1987).
The decoherence of the electromagnetic field coupling resonantly to a
two-level system~(Anastopoulos et al. 2000) and applications of two-level
decoherence to qubit systems~(Shiokawa et al. 2004) are also investigated
in the literature.

The pattern for the two-optical beams arising from phase-randomizing process
(McAlister et al. 1997), is in the same structure of the
two-by-two matrix discussed in this paper.  As for the decoherence in
the rest of the universe introduced in this work, the system becomes
more coherent as the time-variable increases.  Although this
"recoherence" process was considered earlier in the
literature~(Anglin et al. 1996), it is premature to expect a two-state
system to gain coherence in the real world.  It is thus very safe to say
that the second Minkowskian space introduced in this paper remains
in Feynman's rest of the universe.

However, this does not prevent us from constructing a physical system
analogous to the decoherent system coupled to a recoherent system.

\section*{Concluding Remarks}
In this paper, we have organized ray and polarization optics using
the language of the Lorentz group.  The Lorentz group has two-by-two
and four-by-four representations.  Both are useful in optics, and
they allow us to gain a unified view of various aspects of optics.
\par
In addition, it was noted that the mathematics applicable to ray
and polarization optics is directly applicable to the internal
space-time symmetries of elementary particles.  Optical systems
are favorable in the sense that each mathematical operation has a
counterpart that can be performed in optics laboratories.

\par
The Lorentz group is also the basic mathematical language for Einstein's
special relativity.  Currently, this group serves useful purposes
in many other branches of physics, including optical sciences.  In
recent years, the Lorentz group served as the underlying language
for squeezed states of light.  It was Dirac who first observed that
the Lorentz boost is a squeeze transformation~(Dirac 1949)
and constructed representations of the Lorentz
group using coupled harmonic oscillators~(Dirac 1963).  Indeed, Dirac's
oscillator representation forms the theoretical foundations of squeezed
states of light~(Yuen 1976, Yurke et al. 1986, Kim et al. 1991).
This aspect of the Lorentz group is by now well known in the optics community,
and the Lorentz group is one of the theoretical tools in quantum optics.
\par

The squeezed state is not the only branch of optics requiring the
Lorentz group.  It can well be applied to Fourier optics~(Bacry et al. 1981)
while its geometry has proven to be useful in designing three dimensional
non-imaging concentrators~(Gutierrez et al. 1996).
Para-axial wave optics~(Sudarshan et al.1983, Makunda et al. 1983)
and Wavelets~(Han et al. 1995) are also known to be representations of this
group.  It is also the underlying language for reflections and refractions
~(Pellat-Finet et al. 1992).

\par

\end{document}